\documentclass{rspublic}
\usepackage{amsmath}
\usepackage{amsfonts}
\usepackage{amssymb}
\usepackage[Symbol]{upgreek}
\usepackage[nointegrals]{wasysym}
\usepackage{mathrsfs}
\usepackage{graphicx}
\newcommand{\real}{{\mathbb R}}
\newcommand{\bfu}{{\bf u}}
\newcommand{\bfx}{{\bf x}}
\newcommand{\bfk}{{\bf k}}
\newcommand{\bfs}{{\bf s}}
\newcommand{\rr}{{{\mathbb R}^3}}
\newcommand{\Upo}{{\Omega}}
\newcommand{\bdy}{{\partial \Omega}}

\newcommand{\bfv}{{\bf v}}
\begin{document}
\title[Solution of Euler Equations]{\large Remarks on singular solutions of the Euler equations}
\author[F. Lam]{F. Lam}
\affiliation{ }
\label{firstpage}
\maketitle
\begin{abstract}{Euler Equations; Finite-time Singularity; Incompressibility; Potential Flow; Laplace Equation; Velocity-Vorticity Compatibility; Yudovich Singular Solution; Lebesgue's Space; Singular Integrals; Sobolev Inequality; Sobolev Embedding; H{\"o}lder Inequality; Ladyzhenskaya Inequalities}
We examine the blow-up claims of the incompressible Euler equations for several specific flow-fields, (1) the columnar eddies in the vicinity of stagnation; (2) a quasi-three-dimensional structure for illustrating oscillations and concentrations in shears; (3) a 2-and-1/2D flow. We assert that these claimed finite-time singularities are not genuine. Over the whole space, the potential or the velocity coincides with singularities of harmonic functions. We show that the existence of the potential for unique velocity is necessary but insufficient, as the velocity fields are merely specified up to multiples of the curl of a vector potential, which satisfies a system of degenerated Laplace's equations. We have derived several closed-form formulas in simply-connected domains, as well as orthogonal curvilinear co-ordinates. It follows that steady solutions of Euler's equations are indeterminate. Hence, use of singular potential flow in expounding fluid motions is inept. The formulation of the primitive variables (velocity and pressure) gives rise to non-unique flow fields. In essence, the notion of a finite-time singularity is an ill-defined proposition.

In \ref{yud}, we show that the well-posedness claim (Yudovich, 1995) of planar vorticity in Lebesgue's class has been made on the basis of flawed technicalities. 

Certain inequalities of functional analysis play key roles in the regularity study. In \ref{siq}, we assert that their derivations have not been rigorously formulated. Concretely, applications of Sobolev's inequality (as well as Ladyzhenskaya's inequalities) to non-trivial solenoidal velocity fields are unattainable. There have been claims of singular flow solutions in the technical literature. However, a large majority of the deductions rely on these imprecise functional inequalities and hence suffer from logical inconsistency. 
\end{abstract}
\tableofcontents
\section{Background}
The evolution of inviscid incompressible flows of unit-density in space and in time is governed by the Euler equations
\begin{equation} \label{euler}
	 \partial_t \bfu  + (\bfu \cdot \nabla) \bfu = - \nabla p,\;\;\; \nabla{\cdot}\bfu=0,
\end{equation}
with velocity $\bfu{=}(u,v,w)(\bfx,t),\; t \geq 0$. The assumption is that viscous effects may be weak and hence negligible so that the fluid viscosity of the Navier-Stokes equations is formally reduced to zero. The scalar pressure can be eliminated from the equation, giving rise to the vorticity dynamics 
\begin{equation} \label{vort}
	\partial_t \upomega = (\upomega \cdot \nabla) \bfu  - (\bfu \cdot \nabla )\upomega,
\end{equation}
where $\upomega = \nabla {\times} \bfu$, having components $(\xi,\eta,\zeta)$. The vorticity field inherits the velocity solenoidal property, $\nabla{\cdot}\upomega=0$. 
The initial velocity is smooth and bounded, 
\begin{equation} \label{ic}
 \bfu(\bfx,0)=\bfu_0(\bfx).
\end{equation} 
Hence the starting motions always have finite energy. 

The adjoints of the Euler equations are given by
\begin{equation} \label{euleradj}
	\partial_t \bfu^{\dagger} + (\bfu^{\dagger} {\cdot} \nabla) \bfu^{\dagger} = - \nabla p^{\dagger},\;\;\; \nabla{\cdot}\bfu^{\dagger}=0, 
\end{equation}
where $\bfu^{\dagger}(\bfx,t)$, $0\leq t \leq T<\infty$. In the present note, the adjoint `initial' condition is denoted by
\begin{equation} \label{icadj}
 \bfu^{\dagger}_0(\bfx)=\bfu(\bfx,T). 
\end{equation}
The adjoint momentum describes a dynamics backward in time with the identical non-linearity. Because the inviscid fluid mechanics is formulated in a non-dissipative continuum, the temporal reversal in (\ref{euler}) and (\ref{euleradj}) is expected. We shall not write down the adjoint vorticity ($\upomega^{\dagger}$) whose equation of motion has the form of (\ref{vort}).

To avoid excessive technicality, we restrict our discussions to domains $\real^3$ or finite $\Upo$ with $C^1$ boundary $\bdy$ with zero normal velocity,
$\bfu(\bfx,t) \cdot \vec{n} (\bfx) = 0, \bfx \in \bdy$,
where $\vec{n}$ denotes the outward normal. 
We exclude periodic domains with periodic boundary conditions because the initial value problem of the Euler equations is ill-posed in such a formulation.

In practice, any Euler solution must be kinematically consistent with the incompressibility, $\nabla{\cdot}\bfu=0$, that in turn enforces the following compatibility: 
\begin{equation} \label{cmp}
\Delta \bfu = - \nabla{\times}\upomega.
\end{equation} 

The debate on whether the Euler equations can develop a finite time singularity from smooth initial data of finite energy has a long history.
Advocates of finite-time singularities anticipate that the emergence of singular flow structures, whatever they are, may identify the origin of turbulence, or solve the riddle of intermittency in turbulent flows. With an ambitious theoretical intention, it is proposed that the fluid dynamics may have been inadequately formulated.

Following Leray's procedures of establishing existence of the weak Euler solutions, we get the energy inequality,
\begin{equation} \label{eiq}
\frac{1}{2}\int_{\real^3} |\bfu(\bfx,t)|^2 \rd \bfx \leq \frac{1}{2}\int_{\real^3} |\bfu_0(\bfx)|^2 \rd \bfx,
\end{equation}
which cannot be compatible with blow-up scenarios because, close to a singular time $T_s$, the energy on the left is expected to increase suddenly, possibly beyond bounds, unless a singularity has other meanings. At any rate, any potential blow-up is not a Leray-Hopf weak solution.

Indeed, there have been attempts to view singularities in other contexts. A popular one is to interpret Richardson's cascade as a process which proceeds continuously toward arbitrarily small scales, i.e., singularities form due to some mechanism, and certain quantities become somehow non-differentiable. The confusion here is between the sizes of incompressible fluid elements and their velocities. Within the framework of the continuum, the fluid material cannot be indefinitely small so that a vacuum state is created. In the vacuum state there is no fluid so it is absurd to talk about local dynamics. The cascade effects on fluids which are physical matter, observing the principle of mass conservation. In other words, the supports of elements' velocities do not vanish in general. Recall that differentiation works on infinitesimals. Time is a continuous independent variable; the incompressibility constraint puts an upper bound on the flow speed (e.g. (\ref{cmp})).   While one or more of the velocity components may be locally zero, the vorticity or strains may not.
Ultimately, the regularity of the flow field must be sought in the solutions of the Euler equations. We believe that the controversy about Onsager's conjecture of anomalous energy dissipation can be settled in the same sense. The task of showing initial smooth data remaining differentiable throughout the motion is no longer intractable, given our improved understanding of the non-linearity.

Although the nature of a singular structure lacks a precise definition, numerous attempts have been made in the hope of quantifying flow breakdown. Anticipation is whether singular flows do engender multiplications of length scales, leading to the familiar compositions of turbulence, say at locations far away from solid boundaries. It is acceptable that an uncompromising approach is to concentrate on the interior flow complexity, excluding boundary effects. 

Direct numerical simulation of ideal flow in periodic domains with periodic boundary data are known to be incomplete and misleading, if not false, as the loosely-defined boundary velocities inherit discretisation errors which are propagated and amplified over time-marching. Moreover, we notice that the irregular numerics are detected only for a handful of atypical or tailor-made initial data of multiple strong symmetry. At least, the general applicability of singularity hypothesis is largely discounted. Isn't turbulence abundant and ubiquitous?

Apart from the ill-fated numerical syntheses, analytical solutions which demonstrate singular behaviours are scarce. Given any Euler blow-up, we can always solve the adjoint (\ref{euleradj}) backward, using the data at, or prior to, the blow-up time, practically setting $T=T_s-\varepsilon^+$ in (\ref{icadj}). The questions are: Do we recover the original finite-energy flow $\bfu_0$, right down to time $t=0$? What are the underlying mechanisms, if any, which galvanise the two antithetic possesses, the length-scale proliferation and the small eddy annihilation or redistribution, over the {\it identical} time-span? 

Before we address the fundamental issues raised by blow-ups, at least, we expect that any Euler singular solution is analytically or kinematically self-consistent. In this note, we assess two singularity claims.
\section{Columnar vortices}
Consider the flow field decomposition into a quasi-two-dimensional solution field:
\begin{equation} \label{colvt}
	(\bfu,p)(\bfx,t)=\big( \: u(x,y,t),\;\; v(x,y,t),\;\; z\gamma(x,y,t),\;\;p(x,y,z,t)\: \big),
\end{equation}
see Gibbon, Fokas \& Doering (1999); and Malham (2004). For $z \in \real^1$, vortex system (\ref{colvt}) implies an infinite amount of energy for $t\geq 0$. Nevertheless, the vorticity is calculated as
$\upomega=(z\gamma_y, -z\gamma_x, v_x{-}u_y)$.
Thus, the Euler equations are degenerated into a strain field $\gamma(x,y,t)$ and the planar vorticity $\zeta(x,y,t)$. The mass conservation becomes $u_x+v_y=-\gamma$, and, in particular, the $z$-momentum is transformed into
\begin{equation} \label{zz}
	z\; (\gamma_t + u\gamma_x + v\gamma_y + \gamma^2) = -p_z.
\end{equation}
Thus we have three scenarios:
\begin{enumerate}
	\item First, $z$ is non-zero finite. It is legitimate to differentiate (\ref{zz}) to obtain
	\begin{equation} \label{gam}
		\gamma_t + u\gamma_x + v\gamma_y + \gamma^2+p_{zz}=0.
	\end{equation}
	\item Second, $z=0$. For bounded $\gamma$, the velocity field $\bfu$ is purely planar ($p_z\equiv0$) while the vorticity satisfies the convection dynamics, $\zeta_t + (\bfu{\cdot}\nabla)\zeta =0$.
The solution of this quasi-linear diffusion equation is well-posed and its properties are fully understood. 
	\item At $z \rightarrow \infty$, $z \gamma$ is indeterminant.
\end{enumerate}
Thus any reduction (\ref{colvt}) cannot be independent of $z$ on the real line $-\infty {<} z {<} + \infty$. Practically, our observation limits the flow region and the boundary conditions when solving the non-linear equation for $\gamma$. For instance, the tube-like domain proposed in Ohkitani \& Gibbon (2000), $\Upo: \{((x,y),z) \in (\real/L{\mathbb Z})^2 {\times}\real\}$, has a local discontinuity at the plane $z=0$ at all times and, strictly speaking, is ill-defined at $z \rightarrow \pm \infty$; their equation (14) does not hold at these aberrant regions, and the computational results are unconvincing. Likewise, the claims of emerging singular events made by Mulungye, Locas \& Bustamante (2016) must be a consequence of computing artefacts embedded in their quasi-analytic scheme which rearranges the {\it a priori} unboundedness into an excessive computation.

One possible remedy is to work in a finite tube, (say) $0<\epsilon^+\leq z \leq L$, possibly, with periodicity in the other two directions. Yet the initial and boundary conditions at the planes $z=\epsilon^+$ and $z=L$ must be supplemented, and, necessarily, the resulting flow is fully three-dimensional. In addition, the finite tube eschews the absurdity of infinite energy.  
Next, we notice that the di-vorticity, $\nabla{\times}\upomega$, 
\begin{equation*}
\big(\: -u_{yy}+v_{xy}+\gamma_x, \;\;\; u_{xy}-v_{xx}+\gamma_y, \;\;\; -z \gamma_{xx}\:\big),
\end{equation*}
leads to the Poisson equations (cf. (\ref{cmp})),
\begin{equation}
u_{xx}=-v_{xy}-\gamma_x,\;\;\;\;\; v_{yy}=-u_{xy}-\gamma_y, \;\;\;\;\; \mbox{and} \;\;\;\;\; \Delta w=z(\gamma_{xx}-\gamma_{yy}),
\end{equation}
at every given instant $t$. The first two equations are just the continuity as verified by direct integrations. On the plane $z=0$, the third becomes Laplace's equation which has infinitely many non-trivial solutions in $(x,y,z)$. Because of their arbitrariness, these solutions cannot be matched with those obtained from solving (\ref{gam}), particularly in the region $|z| < \epsilon^+$. 

By the same token, there are two discontinuous kinks in Stuart's (3-1), one at the $y{-}z$ plane ($x=0$), and the other at the $x{-}y$ plane (Stuart 1987); the quasi-$2D$ field, his (3-2) to (3-5), is more obscure than the case of one discontinuity as it cannot be completely void of the $x$ or $z$ dependence, taking into account the irregularity at $x,z \rightarrow \pm \infty$. 

In lower dimensions, the same principles apply. For example, the transform of Childress {\it et al.} (1989) makes sense for similitudes, $(u,v)=(\phi(x,t),-y\phi_x(x,t))$, only in finite domains because their equations (1.2$a$) and (1.4) must have been obtained from the condition
\begin{equation*}
y \; \big(\: \phi_{xxt} + \phi \phi_{xxx} - \phi_x\phi_{xx} - \nu \phi_{xxxx}\: \big)=0.
\end{equation*}
Thus the governing equations for $\phi$ break down at two boundaries, the lines $y=0$ and $y=\infty$. In the first instance, the continuity fails as $\phi_x \neq 0$.

For quasi-two-dimensional flows in the cylindrical co-ordinates $(r,\theta,z)$, similar arguments work. Consider the paper of Gibbon {\it et al.} (2003). In view of rotational symmetry, the flow field is assumed to be $(u,v,w,p)=(u,0,z\gamma,p)(r,t)$. Then the momenta are given by
\begin{equation} \label{rmon}
u_t+uu_r=-p_r,\;\;\;\;\;\; z \: \big( \: \gamma_t + u \gamma_r + \gamma^2 \: \big)=0.
\end{equation}
The finite-time solutions (see their equations (4)-(7)) were found from the first equation and the identity, $ \gamma_t + u \gamma_r + \gamma^2 = 0$,
which was fixed regardless of $z$. In fact, its existence must depend on where a plane section is cut across the $z$-direction $[-\infty, +\infty]$. On the plane $z=0$, the non-trivial dynamics of (\ref{rmon}) is the $u$-component, while $\gamma$ becomes an unspecified quantity but must be finite to avoid an indeterminacy. This planar flow cannot be incompressible unless $u=u(t)$. Lastly, the second equation remains essentially unchanged under a translation $z \rightarrow z+c$ for constant $c$, our assertion at $z=0$ applies to any plane cut on the $z$-axis.  

Lin was the first to show that the Navier-Stokes dynamics admits a class of solutions of the form 
\begin{equation} \label{lin}
 \bfu(\bfx,t)=\big(\: u_1,\;\;u_2+y u_4+zu_5,\;\;u_3+y u_6+zu_7 \:\big)(\bfx,t),
\end{equation}
assuming that the gauge pressure has been fixed (Lin 1958). This general expression is a formal reduction. As a matter of fact, Lin has emphasised the importance of appropriate boundary conditions, which are critical in establishing proper solutions of (\ref{lin}). There are considerable leeways to explicitly specify the seven unknowns. Curiously, he has not given any specified domain or any form of boundary data, probably recognising the role played by the co-ordinates $y,z$ as {\it algebraic factors}. With hindsight, it is instructive to recall his remarks on hydrodynamics: 

\begin{quote}
In the ideal case of zero viscosity (and zero magnetic diffusivity), the field may be expected to become infinite as the boundary is approached. It is to be expected
that this tendency to increase will be counteracted by (both) the hydrodynamic
(and the magnetic) diffusive effects.
\end{quote}

The cases discussed in the present section exemplify the singular behaviour at the boundaries located at infinity, for instance, $\pm z \rightarrow \infty$, since all the columnar flows are derived from expression (\ref{lin}). The possibility of an unjustified blow-up is exactly what Lin had in mind $60$ years ago: to avoid the infinite-energy scenario implicated by the co-ordinates system. Evidently, the authors exploiting irregular characters of vortices (\ref{colvt}) were unaware of the subtlety. The claims of finite-time singularities in the vicinity of stagnation points have been made on ill-judged premises, rooted in misinterpretations of Lin's class. 
\section{Shear oscillation and concentration}
Bardos \& Titi (2010) consider a smooth solenoidal flow-field, 
\begin{equation} \label{dm}
\bfu(\bfx,t)=(u,v,w)=\big(\: u(y),\;\; 0, \;\; w(x-tu(y),y) \:\big),
\end{equation}
where $z$ dependence is absent. For purposes of simplification rather than any physical relevance, it is assumed that the pressure may be ignored. As a counter-example, study of this particular flow is expected to show that the Euler equations  are ill-posed in H\"older space $C^{0,\alpha},\: 0{<}\alpha{<}1$. In the simplifying setting, the Euler equations are reduced to
\begin{equation} \label{mo}
w_t+uw_x=0,
\end{equation}
as the other momentum equations are identically satisfied. At every given time $t$, the vorticity and di-vorticity are found to be
\begin{equation} \label{dmvt}
\upomega=\big(\: -t u_y w_x, \;\; -w_x, \;\; -u_y \:\big)(x,y,t),
\end{equation}
and,
\begin{equation} \label{divt}
\nabla{\times}\upomega=\big(\: -u_{yy}, \;\; 0, \;\; -w_{xx}+tu_{yy}w_x-t^2(u_y w_x)^2 \:\big)(x,y,t),
\end{equation}
respectively. Furthermore, we confirm that the vorticity is indeed solenoidal. However, we realise an inconsistency. Applying the continuity constraint (\ref{cmp}), we obtain
\begin{equation} \label{vwp}
v_{xx}+v_{yy}=0,\;\;\;\;\;\; w_{xx}+w_{yy} = t^2 (u_y\; w_x)^2=0,
\end{equation}
where we have used identity, $w_{yy}=-t u_{yy} w_x$, in the last expression. We confirm that the $u$-component reduces to an equality, $u_{yy}$.
In periodic domains, there exist infinitely many non-zero harmonic functions for component $v=v(\bfx)$; the choice of $v\equiv0$ is not entirely incorrect but rather unmotivated. At any instant $t>0$, the first possibility in the $w$-Laplace, $u_y=0$, is incompatible with the assumed (non-zero) $u$ in (\ref{dm}). For the momentum (\ref{mo}) to be unambiguous, the second possibility implies that $w$ must be a steady function of $y$. 

We choose an explicit example to substantiate our discussion. The following  velocity contains finite energy
\begin{equation} \label{osc}
\bfu(x,y,t)=\big(\: \sin(2\pi m y),\;\; 0, \;\; \sin(2\pi m \varphi ) \:\big),
\end{equation}
where symbol $\varphi(x,y,t)=x-t \sin(2\pi m y )$ denotes a characteristic at every fixed $y$, and $m$ a non-zero integer. To be specific, the domain is taken to have a period of $2\pi$ satisfying the `no-slip' condition. Imposing the boundary data is important in order to avoid periodic boundary conditions which are time-dependent, and certainly cannot be constant or zero. As intended, the vorticity is smooth in $(x,y,t)$
\begin{equation*}
\upomega=-\big(\: t4m^2 \pi^2 \cos(2\pi m \varphi)\cos(2\pi m y),\;\;2\pi m \cos(2\pi m \varphi),\;\; 2\pi m \cos(2\pi m y)\: \big),
\end{equation*}
where, essentially,  the $\xi$-component is proportional to $(2\pi m)^2t$. This analysis is often regarded as a demonstration that certain Euler solutions, not necessarily singular, become arbitrarily large in time ($t \rightarrow \infty$) or with increasing circular frequency (large values of $m$). In fact, the law of mass conservation  (\ref{vwp}) {\it per se} dictates either the component $\xi$ must be independent of $y$ or $\eta$ simply vanishes, i.e., the crippled characteristic $\varphi{=}\pm (2k{+}1)/(4m)$ ($k{=}0,1,2,{\cdots}$) which renders velocity $w$ to $\pm 1$. 
 
An elementary velocity, $u{=}\sin y,v{=}0,w{=}1/( a^2{+}\cos^2(x{-}t\sin y))$, hardly says anything new about the Euler analytical properties. Simply put, the $w$-component is contracted to $1/(a^2{+}1)$ by the continuity.
 
The function (\ref{dm}) was originally introduced to investigate if the Euler solutions maintain the long-time well-posedness in the limit of vanishing viscosity $\nu \rightarrow 0$, see DiPerna \& Majda (1987),
\begin{equation*}
\bfu(\bfx,t)=\big(\: u(y,{y}/{\nu}),\;\; 0, \;\; w\big(x-tu(y,{y}/{\nu}),\:y,\: {y}/{\nu}\big) \:\big).
\end{equation*}
Our arguments still apply, provided the stretched $z$-domain, $y/\nu$, remains periodic with (normalised) period unity. Specifically, the $v$-solutions in this co-ordinates' direction may consist of oscillating harmonic functions, depending on our preference. Regrettably, the proposed velocity (\ref{dm}) cannot be a solution of the Euler equations since it is incompatible with the kinematics of incompressible flows. The exception is the less interesting case $\bfu=(C_1,0,C_2)$ for some constants $C_1$ and $C_2$. 
\section{$2\frac{1}{2}D$ Axi-symmetric flows}
\subsection*{Approximations}
Consider the Euler equations (\ref{euler}) in the cylindrical polar co-ordinates $(r,\theta,z)$: 
\begin{equation} \label{eucp}
\begin{split}
		\frac{\partial u_r}{\partial t} + u \frac{\partial u}{\partial r} + \frac{v}{r} \frac{\partial u}{\partial \theta} + w \frac{\partial u}{\partial z}- \frac{v^2}{r} & =-\frac{1}{\rho}\frac{\partial p}{\partial r},\\
		\frac{\partial u_{\theta}}{\partial t} + u \frac{\partial v}{\partial r} + \frac{v}{r} \frac{\partial v}{\partial \theta} + w \frac{\partial v}{\partial z}+ \frac{u\:v}{r}& =-\frac{1}{\rho \: r}\frac{\partial p}{\partial \theta},\\
		\frac{\partial u_z}{\partial t } + u \frac{\partial w}{\partial r} + \frac{v}{r} \frac{\partial w}{\partial \theta} + w \frac{\partial w}{\partial z} & =-\frac{1}{\rho}\frac{\partial p}{\partial z}, 
		\end{split}
\end{equation}
and the continuity,
\begin{equation} \label{dvcp}
\frac{\partial u}{\partial r}+\frac{u}{r}+\frac{1}{r}\frac{\partial v}{\partial \theta} + \frac{\partial w}{\partial z}=0,
\end{equation}
where the components of $\bfu(r,z,t)=(u,v,w)(r,z,t)$.

These equations may be reduced to simpler forms by assuming $\theta$-symmetry. Fluid motions described by the resulting equations depend on co-ordinates $r$ and $z$ only. Such flow-fields are called $2\frac{1}{2}D$ flows, see \S 2.3.3 of Majda \& Bertozzi (2002). Dropping all $\theta$-dependent terms in dynamics (\ref{eucp}), we get the modified equations
\begin{equation} \label{eu0}
\begin{split}
{\partial_t u}  + u \frac{\partial u}{\partial r} + w \frac{\partial u}{\partial z} - \frac{v^2}{r} & = -\frac{1}{\rho}\frac{\partial p}{\partial r},\\
{\partial_t v} + u \frac{\partial v}{\partial r} + w \frac{\partial v}{\partial z} + \frac{u\:v}{r} & = 0,\\
{\partial_t w} + u \frac{\partial w}{\partial r} + w \frac{\partial w}{\partial z} & = - \frac{1}{\rho}\frac{\partial p}{\partial z}.
\end{split}
\end{equation}
The continuity now reads
\begin{equation} \label{icp}
\nabla{\cdot}{\bfu} = \frac{\partial u} {\partial r} + \frac{u}{r} + \frac{\partial w} {\partial z} =0.
\end{equation} 
Denote vorticity $\upomega=(\xi,\eta,\zeta)$. The components in $(r,\theta,z)$ are given by
\begin{equation} \label{vt0}
\xi = - \frac{\partial v}{\partial z},\;\;\; 
\eta = \frac{\partial u}{\partial z} - \frac{\partial w}{\partial r},\;\;\;\mbox{and},\;\;\; \zeta = \frac{1}{r}\frac{\partial (rv)}{\partial r},
\end{equation}
respectively. The continuity (\ref{icp}) is re-written in terms of a stream function,
\begin{equation} \label{phi}
u=-\frac{\partial \phi}{\partial z},\;\;\;w=\frac{1}{r}\frac{\partial (r\phi)}{\partial r},
\end{equation}
so that $\phi$ is driven by component $\eta$ alone,
\begin{equation} \label{smf}
\frac{\partial^2 \phi}{\partial r^2} + \frac{1}{r} \frac{\partial \phi}{\partial r} + \frac{\partial^2 \phi}{\partial z^2} - \frac{\phi}{r^2} = - \eta.
\end{equation}
The vorticity evolution is governed by
\begin{equation} \label{vdy}
\begin{split}
{\partial_t \xi} + u \frac{\partial \xi}{\partial r} + w \frac{\partial \xi}{\partial z} - \frac{\partial u }{\partial r} \xi& = \frac{\partial u}{\partial z} \Big( \frac{\partial v }{\partial r} + \frac{v}{r}\Big),\\
{\partial_t \eta} + \frac{\partial (u \eta)}{\partial r} + \frac{\partial (w \eta)}{\partial z} & = \frac{2v}{r}  \frac{\partial v}{\partial z}, \\
{\partial_t \zeta} + u \frac{\partial \zeta}{\partial r} + w \frac{\partial \zeta}{\partial z} - \frac{\partial w }{\partial z} \zeta& = - \frac{\partial v}{\partial z} \frac{\partial w }{\partial z}.
\end{split}
\end{equation}
\subsection*{Self-similar solution}
Elgindi (2019) dealt with the unbounded domain,
\begin{equation*}
0 \leq r < \infty, \;\;\; 0 \leq z < \infty,
\end{equation*}
with an odd symmetry of $\eta(r,z)=-\eta(r,-z)$ in a curvilinear co-ordinates system ($s, \varphi$), where 
\begin{equation} \label{nc1}
s = (r^2 + z^2)^{\alpha/2},\;\;\;\vartheta = \tan^{-1}\Big(\frac{z}{r}\Big),
\end{equation} 
for some small positive constant $\alpha \in (0,1)$. The stream-function now reads 
\begin{equation} \label{nc2}
\phi(r,z)= (r^2+z^2)^{1/2} \;\psi(s,\vartheta).
\end{equation}
In the case of no swirl ($v=0$), equation (\ref{smf}) is transformed into
\begin{equation} \label{el2p6}
L(\psi)=\alpha^2 s^2 \frac{\partial^2 \psi}{\partial s^2} + \alpha(\alpha+5)s\frac{\partial \psi}{\partial s} + \frac{\partial^2 \psi}{\partial \vartheta^2} - \frac{\partial}{\partial \vartheta} (\tan\vartheta \: \psi) + 6 \psi= - \omega,
\end{equation} 
with homogeneous conditions, 
\begin{equation} \label{nc3}
\psi(s,0)=\psi(s,\pi/2)=0, \;\;\; s \in [0,\infty),
\end{equation}
and, decay at infinity,
\begin{equation*}
\psi(s,\vartheta) \rightarrow 0,\;\;\;\mbox{as}\;\;\; s \rightarrow \infty.
\end{equation*}
Likewise, components in (\ref{vdy}) can also be re-defined. We write $\eta(r,z)=\omega(s,\vartheta)$.

We postulate that solution to (\ref{el2p6}) is in the form of
\begin{equation}
\psi(s,\vartheta)=\psi_p(s,\vartheta)+A \psi_c(s,\vartheta),
\end{equation}
for arbitrary scale $A$, such that
\begin{equation} \label{ops}
L(\psi_p)=-\omega, \;\;\; L(\psi_c)=(\tilde{L}+\tilde{M})(\psi_c)=0,
\end{equation}
where $\tilde{L}=\alpha^2 s^2 \partial_{ss} {+} \alpha(\alpha+5)s \partial_s$, and,  $\tilde{M}=\partial_{\vartheta \vartheta} {-} \tan\vartheta \partial_{\vartheta} {-} (\sec^2\vartheta{-}6)$. Elgindi studied the first equation. The particular form of the second suggests that we seek $\psi_c$ in a separable form:
\begin{equation}
\psi_c=\varphi(\vartheta)(\varpi(s)+\varpi_f(s)),\;\;\; \vartheta \in [0,\pi/2],
\end{equation}
where $\varpi_f$ is an $L^2$-function, $\varpi_f(0)=0$, and $\varpi_f \rightarrow 0$ as $s \rightarrow \infty$. We have two examples: the first purely decays at $s \rightarrow \infty$, 
\begin{equation}
\varpi_f=s \exp(-s^2), \;\;\; \|\varpi_f\|^2_{L^2}=\frac{\sqrt{\pi/2}}{8},
\end{equation}
and, the next one oscillates:
\begin{equation}
\varpi_f=\sin(2\pi s) \exp(-s), \;\;\; \|\varpi_f\|^2_{L^2}=\frac{\pi/2}{1+\pi^2}.
\end{equation}
We write down the explicit formula, 
\begin{equation*}
L(\psi_c(s,\vartheta))=\varphi(\vartheta) \tilde{L}(\varpi(s)+\varpi_f(s)) + (\varpi(s)+\varpi_f(s))\tilde{M}(\varphi(\vartheta)),
\end{equation*}
where, now, $\tilde{L}$ and $\tilde{M}$ are ordinary operators.

\begin{figure}[ht] \centering
  {\includegraphics[keepaspectratio,height=14cm,width=12cm]{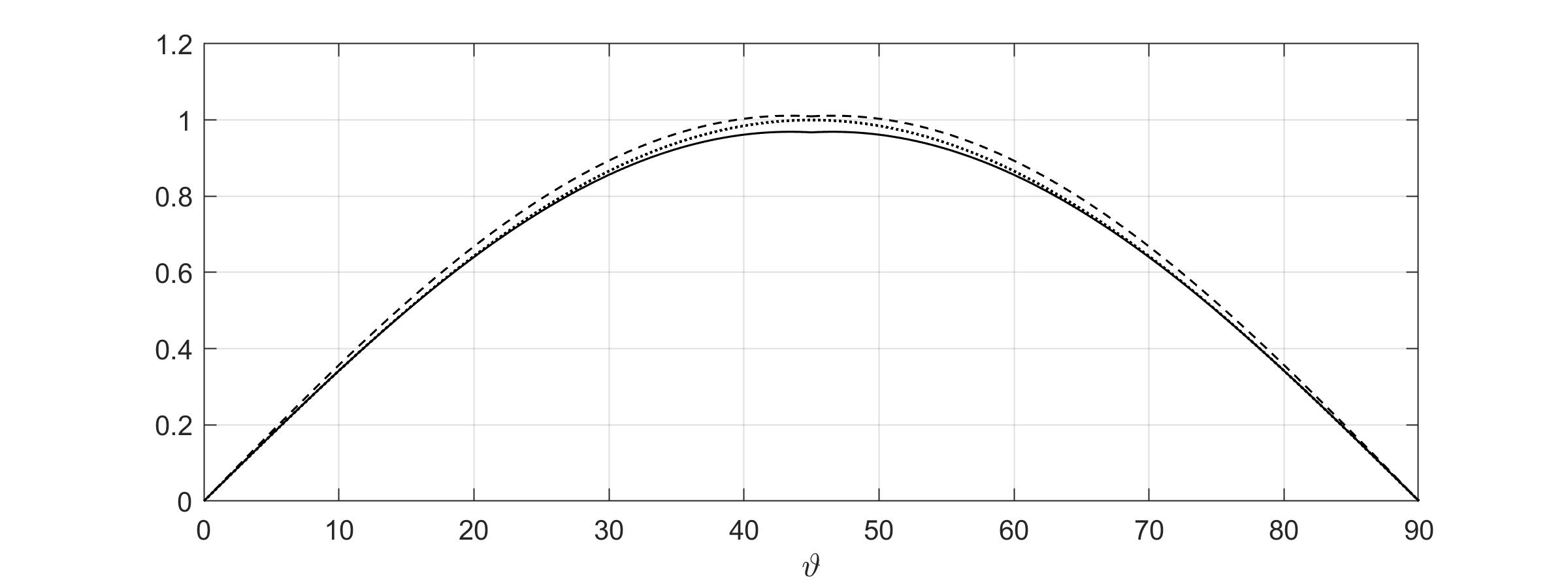}}
 \caption{A solution $\varphi$ of (\ref{fn2}) as perturbations to $\sin(2\vartheta)$ (dotted-line). Boundary value $\beta=1/4$. The calculated result is multiplied by $0.46$ and $0.48$ respectively.} \label{fig1} 
\end{figure}

Let $\beta=25\alpha^2/4$. Our key step is to decompose operator $\tilde{M}$ into
\begin{equation} \label{opm}
\tilde{M}(\varphi)=\tilde{M}^*(\varphi) - \beta \varphi.
\end{equation}
We determine $\varphi$ by solving
\begin{equation} \label{fn2}
\tilde{M}^*(\varphi)=\varphi'' - \tan\vartheta \varphi' -(\sec^2\vartheta-6-\beta)\varphi=0,
\end{equation}
with specific boundary data, $\varphi(0)=0$, $\varphi'(\pi/4)=-\beta$. For every given $\alpha$, we seek the even solution about $\vartheta=\pi/4$. The existence of $\varphi$ follows well-established theory of ordinary differential equations. Thus we may view scaled $\varphi$ as suitable $L^2$-perturbations, see figure~\ref{fig1}.

Once $\varphi$ is known, regular functions operated by $\tilde{L}$ are non-vanishing as
\begin{equation} \label{opl}
\tilde{L}(\varpi)-\beta \varpi = \alpha^2 s^2 \varpi'' + \alpha(\alpha+5) s \varpi' -\beta \varpi = \beta \varpi_f - \tilde{L}(\varpi_f).
\end{equation}
For the homogeneous equation, we have the characteristic polynomial
\begin{equation}
a^2 \lambda (\lambda-1) + a(a+5) \lambda - \beta=0,
\end{equation} 
and its two solutions, $\lambda_+ (>0)$ and $\lambda_- (<)$, are
\begin{equation}
\frac{5}{2\alpha} \big( -1 \pm \sqrt{1+\alpha^2} \big),
\end{equation} 
respectively. To $O(\alpha^4)$, we have 
$\lambda_+ \approx {5}/{4} \alpha,\;\lambda_- \approx -( {5}/{\alpha} + 5 \alpha/4 )$.

It is straightforward to derive Green's function
\begin{equation*} 
	G(s,q)= -\frac{\alpha}{5\sqrt{1+\alpha^2}}\begin{cases}
	\; q^{(\lambda_- + 3/2)}\; s^{\lambda_+}, & 0 \leq s \leq q, \\
	\; s^{\lambda_-}\; q^{(\lambda_+ + 3/2)}, & q \leq s < \infty.
	\end{cases}
\end{equation*}
Thus
\begin{equation}
\varpi(s) = \int_0^{\infty} G(s,q) \big(\beta \varpi_f-\tilde{L}(\varpi_f\big) (q) \;\rd q.
\end{equation}
At every given time prior to blow-up, energy calculated by solution $\psi$ of (\ref{el2p6}) is non-unique and possibly unbounded, being independent of vorticity $\omega$.
\subsection*{Further comments}
Remark 1: The equation governing stream-function (\ref{el2p6}) has a specific form and, apparently, has been designed in view of curvilinear system (\ref{nc1}). In particular, it takes the identical form if the swirl is non-zero ($v\neq 0$). It follows that the indeterminacy of the velocity field can be adapted to challenge the stability hypothesis of self-similar solutions (Elgindi {\it et al.}, 2019). Moreover, elementary trigonometric functions, $\sin(2 \vartheta)$ and $\sin\vartheta \cos^2\vartheta$, play no important role in fluid motions. Data (\ref{nc3}) are tailor-made for certain boundary terms to vanish. Delicate perturbations other than those from $\tilde{M}^*(\varphi)=0$ may be defined in general function spaces.

Remark 2: If $\alpha {\rightarrow} 0$ formerly, hypotenuse $s$ is becoming independent of $r$ and $z$. Effectively, direction $\vartheta$ is rendered insignificant. If a velocity field in $(r,\theta)$ is finite at an instant $t$, so is $\phi$. In turn, $\psi$ is imprecise according to (\ref{nc2}). In the limiting process, the liaison between (\ref{nc1}) and (\ref{nc2}) is not necessarily consistent for bounded $\eta(r,z)$ in the vicinity of the origin.

Remark 3: The Euler equations, the continuity law as well as the inviscid vorticity dynamics {\it are invariant} with respect to any shift in the $z$ direction. Recall the fact that the cylindrical Euler equations (\ref{eucp}) and (\ref{dvcp}) are already in their canonical forms. We may shift $z$ to $z{-}z_0$ in (\ref{nc1}), (\ref{nc2}) and (\ref{nc3}) so as to define the origin. Suppose there is a regular solution of $\psi$, and $\psi(s(r,z),\vartheta)=\psi(s(r,z{-}z_0),\vartheta)$. However, the corresponding $\phi(r,z)$ is seen to be scale-variant by the choice of $z_0$ (at fixed $r$ and time $t$). The moral is that algebraic formula (\ref{nc2}) does not respect the invariance principle. 
\subsection*{Numerical simulations}
To avoid the infinite domain on the $z$ co-ordinate, numerical solutions may have to be carried out in truncated domains. Consider a pill-box ($0 \leq \theta < 2 \pi$):
\begin{equation*}
\Upo_B: \;\;\; 0 \leq r \leq 1, \;\;\; 0 \leq z \leq 1, 
\end{equation*}
and, assume periodic flows.

The dynamic equations, (\ref{eu0})-(\ref{vdy}), are solved with smooth initial conditions,
\begin{equation} \label{ic0}
\bfu_0(r,z) \in C_c^{\infty}(\Upo_B).
\end{equation}
The no-flow on the lateral wall requires
\begin{equation} \label{nf0}
u(r=1,z,t\geq0)=0.
\end{equation}
Imposing (explicit) periodic boundary conditions is crucial in the present geometry. The solutions have to be identical on the lid and the bottom
\begin{equation} \label{pc0}
\bfu(r,z=0,t\geq0)=\bfu(r,z=1,t\geq0).
\end{equation}
There is no {\it a priori} vorticity boundary condition.
Any fluid motion $(\upomega,\bfu,p)$ in $\Upo$ is completely specified. The general approach to determine the pressure $p$ is to solve a Poisson's equation with appropriate Neumann data on $r=1$.

First, we assert that the following velocity-vorticity field furnishes a class of exact solutions:
\begin{equation} \label{uv0}
\begin{split}
\bar{u}&=0, \;\;\; \bar{v}=g(r), \;\;\; \bar{w}=A h(t) - B r;\\
\bar{\xi}&=0,\;\;\; \bar{\eta}=B,\;\;\; \bar{\zeta}=\bar{\zeta}(r).
\end{split}
\end{equation}
where $A$ and $B$ are arbitrary finite constant; $h(t)$ any function of time $t$. The gradient, $\partial_z p = q(t)$, is obtained from (\ref{eu0}). Second, $\bar{\zeta}$ can be computed from (\ref{vt0}). We are free to choose any regular $g$ as long as the enstrophy is finite. From the $r$-momentum, the compatible centrifugal force, $\partial_r p$, can be found. The class of flows with zero-swirl is a special case, $g(r) \equiv 0$. 

Let $\tilde{\bfu}$ and $\tilde{\upomega}$ be the numerical computations of Chen \& Hou (2022) over interval $t \in [0,\tilde{t}^*)$, where $\tilde{t}^*$ denotes their blow-up time. Let $\tilde{E}(t^+)$ be the energy at $t^+$ for $0 < t^+ \ll \tilde{t}^*$. Let $\delta R$ and $\delta t$ denote the (non-dimensional) discretisation errors in space and in time respectively. Similarly, let $\delta I$ denote the interpolation errors. Then the following bounds exist
\begin{equation}
\Delta_r < \min( |\delta R(t)|, |\delta I(t)|) ;\;\;\; \Delta_t < \min ( |\delta t(t)| ),
\end{equation}
for $0 < t \leq t^+ $. Next, we denote the truncation errors (if any) by $\delta N(t)$ if the solutions are approximated by expansions of eigen-functions.  We extend $\Delta$ to include these additional errors,
\begin{equation}
\Delta_r < \min( |\delta R(t)|, |\delta I(t)|, |\delta N(t)| ).
\end{equation}

Set $\Delta = \max ( \Delta_r, \Delta_t ) (> 0)$. For $n=1,2,3,\cdots$, we choose sequences 
\begin{equation} \label{d1}
|A_n|< \Delta, \;\;\; |B_n| < \Delta,
\end{equation} 
and,
\begin{equation} \label{d2}
|\zeta_n| < \Delta, \;\;\; |g_n| < \Delta, \;\;\; (0 \leq r \leq 1)
\end{equation}
and,
\begin{equation} \label{hn}
h_n (t-t_0) = \frac{\sin{(n (t-t_0))}}{\pi (t-t_0)},\;\;\; 0 \leq t_0 < t <\Delta.
\end{equation} 

For every initial data, $\tilde\bfu_0(r,z)$ or $\tilde\upomega_0(r,z)$, we add to their numerically-computed solutions to obtain admissible approximations:
\begin{equation} \label{uv}
\bfu \approx\bar{\bfu}+\tilde{\bfu},\;\;\; \upomega \approx \bar{\upomega}+\tilde{\upomega},\;\;\; t \in (0,t^+].
\end{equation}
By the well-known properties of functions given in (\ref{hn}), there exist instants, $t < \Delta$ so that $\bar{E}(t) \gg \tilde{E}(t^+)$. Now consider perturbations to our solutions $\bar{\bfu}$ and $\bar{\upomega}$ by re-defining $\Delta$ and $t_0$. Set $\Delta_{\varepsilon}=\Delta-\varepsilon_{\Delta}$, $\varepsilon_{\Delta} > 0$, and, $t_0=t_{\varepsilon} + \varepsilon_t$, $\varepsilon_t > 0$. Moreover, $\varepsilon_{\Delta} \sim O(0^+)$, $\varepsilon_t \sim O(0^+)$. We replace bound $\Delta$ in (\ref{d1}) and (\ref{d2}) by $\Delta_{\varepsilon}$. Likewise, we modify $h_n$ such that
\begin{equation} \label{hn2}
h_n (t-t_{\varepsilon}) = \frac{\sin{(n (t-t_{\varepsilon}))}}{\pi (t-t_{\varepsilon})},\;\;\; 0 \leq t_{\varepsilon} < t <\Delta_{\varepsilon}.
\end{equation}
It follows that, on the wall $r=1$ for $0<z<1$, there exist numerous temporal locations when the local $\bar{E}(t_{\varepsilon}) \ll \tilde{E}(t^+)$ due to the oscillatory characters of function $h_n$ as $n \rightarrow \infty$. 

Hence solutions (\ref{uv}) must be {\it numerically} unstable.
We have also established a corollary that these solutions are non-unique. The existence of the degenerated flows (\ref{uv0}) asserts that the inviscid axi-symmetric formulation in geometry $\Upo_B$ is ill-posed in the sense of Hadamard (Book I of Hadamard, 1923).
\subsection*{Boussinesq equations}
A model describing the motion of fluids with stratified density ($\varrho$) under the influence of gravitational forces is known as the Boussinesq approximations. The domain of interest is a strip of infinite extent, $ - \infty < x < + \infty $, $0 \leq y \leq L_y$. To avoid the difficulties of dealing with the unbounded planar domain, we shall restrict our attention to a periodic box, $0 \leq x \leq L_x$. The equations of motions are given by
\begin{equation} \label{beq}
\begin{split}
\partial_t \omega + u \frac{\partial \omega}{\partial x} + v \frac{\partial \omega}{\partial y} & = - \frac{\partial \varrho}{\partial x}, \\
\partial_t \varrho + u \frac{\partial \varrho}{\partial x} + v \frac{\partial \varrho}{\partial y} & = 0,
\end{split}
\end{equation}
where $(u,v,\omega,\varrho)=(u,v,\omega,\varrho)(x,y,t)$. The vorticity $\omega=v_x-u_y$, and, $u_x+v_y=0$. We impose the conditions of
\begin{equation} \label{beq1}
\begin{split}
v(x,y=0,t\geq0)&=v(x,y=L_y,t\geq0)=0,\\
(u,v,\varrho)(0,y,t\geq0)&=(u,v,\varrho)(L_x,y,t\geq0).
\end{split}
\end{equation}
The initial data are assumed to be smooth: 
\begin{equation} \label{beq2}
(u_0,v_0,\varrho_0)(x,y)=(u,v,\varrho)(x,y,t=0).
\end{equation}

The analogous solutions to (\ref{uv0}) are expressed as
\begin{equation} \label{uvr0}
u=a f_1(t) - b f_2(y),\;\;\;v=0,\;\;\;\omega=b\:(\rd f_2/\rd y),\;\;\;\varrho=g_1(y),
\end{equation}
where $a$ and $b$ are finite constants, $f_1(t)$, $f_2(y)$ and $g_1(y)$ arbitrary bounded functions. By analogy to the axi-symmetric case, we conclude that the Boussinesq dynamics (\ref{beq})-(\ref{beq2}) cannot be uniquely defined as well as suffers from numerical instability. 
\subsection*{Discussion}
Remark 1: In the case of flows without swirl ($v=0$), Majda \& Bertozzi (2002) introduced a quantity $\hat{\eta}=\eta/r$. Then the second equation in (\ref{vdy}) is reduced to
\begin{equation*}
r {D \hat{\eta}}/{D t} \equiv r \big( {\partial_t \hat{\eta}} + u {\partial \hat{\eta}}/{\partial r} + w {\partial \hat{\eta}}/{\partial z} \big) = 0,
\end{equation*}
by virtue of the continuity. Consequently, we cannot conclude that $\hat{\eta}$ is conserved along particle trajectories over the $whole \;domain$ $\Upo$ (cf. their equation (2.58) or (4.60)). On the axis $r=0$, the total rate of change, $D \hat{\eta}/Dt$, can be any bounded non-zero function in $(r,z,t)$. As $r \rightarrow \infty$, the total derivative is undefined.

Remark 2: For numerical purposes, Luo \& Hou (2014) chose to work with $\hat{v}{=}v/r$, $\hat{\eta}{=}\eta/r$, and, $\hat{\phi}{=}\phi/r$. Several strong symmetries have been imposed. Their choice of the starting profile concentrated high gradients near the wall region, thus, casting doubts on whether the incompressibility hypothesis remains valid at all times. There was no mention of computed velocities near their potential blow-up ($\hat{T}^*$). Nevertheless, we see that a spurt flow of zero swirl,
$\hat{\phi}_s=A f(t)/2,\; \hat{\eta}_s=0$,
is an exact solution to their [2a]-[2c]. It ensures zero flow on the wall and satisfies the pole and periodicity conditions. Indeed, their axial velocity consists of two parts,
\begin{equation} \label{wv0} 
w(r,z,t) \approx Af(t) + (2 \hat{\phi} + r \partial \hat{\phi}/\partial r)(r,z), \;\;\; t \in [0,\hat{T}^*),
\end{equation}
where function, $Af(t)$, can be chosen arbitrarily ($ f(0)=0$ to be consistent with the initial data). We would encounter absurd scenarios where scale $A$ itself may instigate a flow of excessive energy ($\propto w^2$) which violates the continuity. Thus their claimed singularity is no more than an estimation inferred from a differential system whose analytical well-posedness is paradoxical.

Remark 3: There exist other non-trivial exact solutions in $z$-periodic domains in addition to (\ref{wv0}). For analogous solutions in domains with multiple periodicity, see Lam (2018). Consider equations (\ref{eu0})-(\ref{vdy}) which include  (1.2)-(1.3) of Chen \& Hou as a subset. We have one irrotational spurt without swirl:
\begin{equation} \label{wv1} 
u=0,\;\;\; w=Af(t),\;\;\;p=p_0 - A \rho z \: (\rd f/\rd t),\;\;\; \phi=A r f(t),
\end{equation}
and one with non-zero $\upomega$:
\begin{equation} \label{wv2} 
\begin{split}
u&=0,\;\;\;v= v(r)=A_1 r g(r),\;\;\; w= A_2 f(t),\;\;\; \phi=A_2 r f(t),\\
p&=p(r,z;t)=p_0 + A^2_1 \rho \int^r \!g^2(r) r\; \rd r - A_2 \rho z \: (\rd f/\rd t),\\
\xi&=\eta=0,\;\;\; \zeta=\zeta(r)=A_1 \big(2g+r\: (\rd g/\rd r) \big).
\end{split}
\end{equation}
Within an infinitesimal interval after motion's onset, any solution of Chen \& Hou (2022), obtained from numerical means or operator approximations or self-similarity construction, has been decimated by superimposing sequences of these spurts. The question is: Do their linear and non-linear stability analyses still make sense in the present context?

Remark 4: There is an equivalent form of the stream function (cf.(\ref{smf})):
\begin{equation*} 
u=-\frac{1}{r}\frac{\partial \phi}{\partial z},\;\;\;w=\frac{1}{r}\frac{\partial \phi}{\partial r},\;\;\;\mbox{and},\;\;\; \frac{\partial^2 \phi}{\partial r^2}-\frac{1}{r}\frac{\partial \phi}{\partial r}+ \frac{\partial^2 \phi}{\partial z^2}=-r \eta.
\end{equation*}
The $\phi$ solution in (\ref{wv1}) or (\ref{wv2}) is now $\propto r^2f(t)$. However, the elliptic equation appears to offset the algebraic proportion for the operators (\ref{ops}). Due to their own auxiliary nature, either form has limitations in describing complete flow-fields.

Remark 5: The presence of the coefficient $1/r$ in the governing equations (\ref{eucp}) is the consequence of co-ordinates's transformation, $x=r \cos\theta, \:y=r \sin\theta$. The factor $1/r$ does not imply any fundamental issues in flow-fields, as it is nothing more than the origin of the Cartesian counterpart. If a curvilinear co-ordinates mapping is bijective, is it imperative to look for singular behaviours near the Cartesian origin? On the other hand, an occurrence of blow-up on smooth boundaries at fixed $(r,\theta)$ has to be an analytical or numerical artifact as, by implication, the whole section or line along the $z$-axis becomes singular (in view of the invariance). Physics sanctions the collapsing topologies; no measurable flows evolve in this manner. 

Remark 6: Because the Euler equations deal with motions of {\it ideal} fluids, observations from laboratory experiments cannot be explained by the inviscid theory, as the detailed mutual vortex interactions among multitudinous spatio-temporal scales are intrinsically viscous. Emergence of a finite-time singularity necessarily involves large velocities; there are no justifications to invalidate the law of mass conservation. Perhaps, there are no inviscid continuum flow-fields which can be treated as 2-and-1/2 dimensions. For a general smooth data, $\bfu_0(\bfx)$, the degeneracy of the full dynamics (\ref{eucp})-(\ref{dvcp}) by azimuthal-symmetry warps the flow-field to such an extent that analysis fails.

\section{Flow indeterminacy by a scalar potential} \label{pflw}
\subsection*{Introduction}

Potential flows in simply connected regions have the following properties
\begin{equation} \label{pf}
	\nabla{\cdot}\bfu=0,\;\;\;\;\; \nabla{\times}\bfu=\upomega=0,
\end{equation}
see, for instance, Lamb (1932); Prandtl \& Tietjens (1934). The circulation over a closed curve ($C$) equals to the integral of the normal component of vorticity taken over any finite surface ($S$) enclosed by curve $C$ (Stokes' theorem on circulation), 
\begin{equation} \label{stk}
	\Gamma=\oint (u \rd x + v \rd y + w\rd z) = \int_C \bfu \cdot \rd {\bf r}= \int_S (\vec{n}\cdot\upomega) \; \rd \bfs=0,
\end{equation}
where the value of the line integral, taken from a starting position $A$ to end position $B$, is path-independent. Note that the quantity, $\bfu \cdot \delta{\bf r}=u\rd x+ v \rd y +w \rd z$, is an exact differential. The line integral, 
\begin{equation} \label{iph3}
	\int_{A}^{B} \bfu \cdot \rd {\bf r} = \phi_B - \phi_A,
\end{equation}
where $\phi$ is a single-valued function of position, known as scalar potential. We call velocity $\bfu$ a conservative vector field. Over infinitesimal displacement $\delta {\bf r}$, we express the resulting variation in
$\bfu \cdot \delta{\bf r} = \delta \phi = \nabla \phi \cdot \delta{\bf r}$, or, $(\bfu -\nabla\phi) \cdot \delta{\bf r} = 0$. The last equality shows that, for arbitrary $\delta{\bf r}$, the potential $\phi$ exists, and its gradient describes the velocity
\begin{equation} \label{ph3}
	\bfu(\bfx) = \nabla \phi(\bfx).
\end{equation}
By the continuity (\ref{pf}), the velocity field is found by solving the Neumann problem
\begin{equation} \label{pnm}
	\Delta \phi = 0, \;\;\;\;\; \frac{\partial \phi}{\partial \vec{n}}\Big|_{\bdy} = \vec{n}\cdot \bfu\big|_{\bdy} \;\;(\;= {\bfu_s}).
\end{equation}
In Green's first identity, (for $\phi$ and $\psi$ being $C^2(\overline{\Upo})$ functions)
\begin{equation} \label{grn1}
	\int_{\Upo} \psi \Delta \phi \; \rd \bfx = \int_{\bdy} \psi \frac{\partial \phi}{\partial \vec{n}} \; \rd \bfs \;- \int_{\Upo} \big(\nabla \phi \cdot \nabla\psi \big) \; \rd \bfx,
\end{equation}
we put $\psi=1$, the first integral on the right defines the compatibility condition,
\begin{equation} \label{pcm}
	\int_{\bdy} (\vec{n}\cdot \bfu )\; \rd \bfs = 0.
\end{equation}
The solution of Neumann problem (\ref{pnm})-(\ref{pcm}) is well-established.

On reflection, the only thing we know about the kinematics is that $\bfu$ is conservative. The reason that we can carry out the line integral (\ref{iph3}) is because we are given the velocity field $\bfu$. What we are solving is the cause-and-effect problem $\nabla \phi=\bfu$. But the converse, $\bfu=\nabla \phi$, is not evident. Let $\uppsi$ be a vector potential which is a single-valued function of position. If the curl of $\nabla{\times}\uppsi$ vanishes everywhere, $\nabla{\times}\uppsi$ is a conservative field. An alternative to (\ref{iph3}) is the following, 
\begin{equation} \label{ph4}
	\int_{A}^{B} \big(\bfu + \nabla{\times}\uppsi \big) \cdot \rd {\bf r} = \phi_B - \phi_A,
\end{equation}
for the identical potential, only if the zero-circulation hypothesis, $\nabla{\times}\nabla{\times}\uppsi=0$, can be justified. By vector identity, $\nabla{\cdot}(\nabla{\times} \uppsi)=0$, we can add multiples of $\nabla{\times}\uppsi$ to $\bfu$ without altering $\phi$'s Neumann problem. Conversely, any computed potential $\phi$ cannot alone differentiate its velocity contents; its gradient describes $\bfu$ as well as $\bfu+\nabla{\times}\uppsi$, as long as latter's boundary condition does not alter $\bfu$'s. The existence of the scalar potential $\phi$ is only necessary to retrieve a unique velocity $\bfu$ but is {\it not sufficient}.

To clarify the matter further, we observe that the solenoidal $\bfu$ itself having zero-circulation must be harmonic. Expanding expression $\nabla{\times}\nabla{\times}\bfu$ leaves
\begin{equation} \label{pnm2}
	\Delta \bfu =-\nabla{\times}\upomega = 0\;\; \mbox{in}\;\; \Upo,\;\;\;\;\; \vec{n}\cdot\bfu|_{\bdy}=\bfu_s.
\end{equation}
This higher-order formulation is a less stringent way to fix velocity $\bfu$ as the di-vorticity simply vanishes. Examples of non-zero vorticity are: (1) constancy $\upomega={\bf c}$; (2) $\upomega=A(yz,\;zx,\;xy)$; (3) $\upomega=(A x,\;B y,\;C z)$, where $A+B+C=0$. Compared to Neumann problem (\ref{pnm}), solutions of (\ref{pnm2}) are definitive and unique.

\subsection*{Potential flow in $\rr$}

One of the key results in the potential theory is that non-trivial potential flows decaying at infinity contain singularities. If $\bfu=\nabla \phi$ is incompressible, irrotational and bounded at infinity, the velocity potential then satisfies
\begin{equation} \label{lp3}
	\Delta \phi(\bfx) = 0,\;\;\; \phi \rightarrow 0\;\;\; \mbox{as}\;\;\; |\bfx|\rightarrow \infty.
\end{equation}
Singular solutions are well-instructed in introduction to fluid mechanics. The fundamental solutions resemble point sources or sinks:
\begin{equation} \label{slp3}
	\phi(\bfx)=\mp \frac{m}{4 \pi r},\;\;\;\;\;(r\neq 0)
\end{equation}
where $r=\sqrt{(x{-}x_0)^2{+}(y{-}y_0)^2{+}(z{-}z_0)^2}$, with the singularity centred at $\bfx_0$. There are known bounds, $|\partial_{x_i} \phi(\bfx)|\leq C |\bfx{-}\bfx_0|^{-2}$, and $|\partial^2_{x_i} \phi(\bfx)|\leq C |\bfx{-}\bfx_0|^{-3}$, for some positive constant $C$. Another singular solution is called a doublet:
\begin{equation} \label{dbt}
	\phi(\bfx)=-\frac{m x}{4 \pi r^3}\;\;\;\;\;(r\neq 0).
\end{equation}

However, these singular solutions do not produce unique velocity fields because $\Delta \phi=0$ is implied in either relation,
\begin{equation} \label{pot3}
	\bfu + \nabla{\times}\uppsi = \nabla \phi,\;\;\;\;\; \bfu = \nabla \phi.
\end{equation}
Vector equation, $\Delta \uppsi=\nabla(\nabla{\cdot}\uppsi)$, enforces irrotationality $\nabla{\times}\uppsi$. Taking curl on this equality, we verify that $\nabla{\times}\uppsi$ is also harmonic
\begin{equation}  \label{psi3}
	\Delta (\nabla{\times}\uppsi)=\nabla{\times}\nabla(\nabla{\cdot}\uppsi)=0,\;\;\;\;\; \nabla{\times}\uppsi \rightarrow 0 \;\;\; \mbox{as} \;\; |\bfx|\rightarrow \infty.
\end{equation}
One of its solutions represents a cluster of doublets:
\begin{equation}
	\nabla{\times}\uppsi=-\frac{m_1}{4 \pi} \; \bigg(\frac{x-x_1}{r_1^3}, \;\; \; \frac{y-y_1}{r_1^3}, \;\;\; \frac{z-z_1}{r_1^3} \bigg),
\end{equation}
where $r_1=\sqrt{(x{-}x_1)^2{+}(y{-}y_1)^2{+}(z{-}z_1)^2}$. For consistency, we validate the prerequisites, $\nabla{\cdot}(\nabla{\times}\uppsi)=0$, and $\nabla{\times}(\nabla{\times}\uppsi)=0$, by direct calculation. In the case (\ref{slp3}), the presence of the $\uppsi$ solution renders the source flow indefinite, for example, the $x$-component becomes
\begin{equation} \label{uss3}
	u(\bfx) = \frac{m}{4 \pi} \frac{(x-x_0)}{r^3} + \frac{m_1}{4 \pi} \frac{(x-x_1)}{r_1^3}.
\end{equation}
This is why considerations of singularity make little sense, as non-trivial source velocity fields are ambiguous in view of a doublet (for example, $\bfx_0=\bfx_1,m=-m_1$). Due to the linearity of the boundary value problem (\ref{psi3}), velocity $\bfu$ in (\ref{pot3}) can be superposed at will by a series of velocities $\sum_i (\nabla{\times}\uppsi)_i$, each corresponds to a doublet cluster $m_i$ located at $\bfx_i$. Indeed, singular potential flows in the whole space have no precise meanings. At least, we must remember the limitation in demonstrating fluid phenomena.

\section{Degenerated vector potentials in simply connected regions} \label{dvps}

Let $\bfu$ be an incompressible flow with given boundary data on $\bdy$. We are particularly interested in whether it can be superposed in the form
\begin{equation} \label{pp}
	\bfu(\bfx) + \nabla {\times} \uppsi(\bfx).
\end{equation}
The property (\ref{pf}) leads to the following equation for the determination of the vector potential $\uppsi$
\begin{equation} 
	\nabla{\times}\nabla{\times}\uppsi=0; \;\;\;\;\; \nabla{\cdot}\uppsi \neq0, \;\;\;\;\; \vec{n}\cdot(\nabla{\times}\uppsi)\big|_{\bdy}=0.
\end{equation}
The reason of specifying the non-solenoidal constraint is to avoid a Laplace-type equation. In order to keep the boundary data of $\bfu$, we enforce the homogeneous boundary value. We must also examine the nature of any modification in the tangential directions. In smooth domains, inclusion of the vector potential must not introduce irregularities. Representation (\ref{pp}) or $\nabla\phi+\nabla{\times}\uppsi$ resembles a Helmholtz's decomposition which states that any velocity field can be decomposed into an irrotational flow and a solenoidal flow. The essence is that the presence of $\uppsi$ is independent of the divergence and curl of $\bfu$. 

Denote $\uppsi(\bfx)=\big(\alpha,\;\beta,\;\gamma\big)(\bfx)$. They satisfy
\begin{equation} \label{eeq}
\left.
\begin{aligned}
	\partial_{yy}\alpha &+\partial_{zz}\alpha -\partial_{xy}\beta-\partial_{xz}\gamma=0,\\
	\partial_{zz}\beta & +\partial_{xx}\beta -\partial_{yz}\gamma-\partial_{yx}\alpha=0,\\
	\partial_{xx}\gamma&+\partial_{yy}\gamma-\partial_{zx}\alpha-\partial_{zy}\beta=0,
	\end{aligned}
\hspace{5mm} \right\}
\end{equation}
in a number of standard domains. The solution of interest is 
\begin{equation*}
	\nabla{\times}\uppsi(\bfx)=\bfu^*(\bfx)=(u^*,\;v^*,\;w^*)(\bfx).
\end{equation*}

\subsection*{The whole space $\rr$} 

The structure of (\ref{eeq}) admits a singular irrotational solution
\begin{equation*}
	\alpha(\bfx)=\beta(\bfx)=\gamma(\bfx)=\frac{1}{\;4 \pi \big(q+d\big)^n\;},\;\;\;q=x+y+z,
\end{equation*}
where $n$ is a natural number, $d$ is finite. A function $f(q)$, behaving like a `Green function' near the origin, is also a solution, for instance, $\alpha=\beta=\gamma=1/\sin q(\bfx)$. The symmetry in the three independent variables suggests that one non-singular field takes the form
\begin{equation*} 
	\uppsi(\bfx)=\big(\; g_0,\;\;g_0,\;\;g_0\;\big), \;\;\; g_0=1/(q^2+d^2).
\end{equation*}
This is still irrotational. We have to break the symmetry, in order to have non-zero curls. The following functions,
\begin{equation} \label{whs}
\begin{split}
	\alpha & = \Big(\log\big(\; (y-b)^2+(z+c)^2\;\big)-\log\big(\; (y+b)^2+(z-c)^2\;\big)\Big)/2 + g_0,\\
	\beta & = \Big(\log\big(\; (z-c)^2+(x+a)^2\;\big)-\log\big(\; (z+c)^2+(x-a)^2\;\big)\Big)/2 + g_0,\\
	\gamma & = g_0,
	\end{split}
\end{equation}
satisfy system (\ref{eeq}) for finite non-zero $a$, $b$, $c$. It follows that
\begin{equation}
\begin{aligned}
	u^*(\bfx) &= \partial_y\gamma-\partial_z\beta=\frac{z+c}{(z+c)^2+(x-a)^2}-\frac{z-c}{(z-c)^2+(x+a)^2},\\
	v^*(\bfx) & = \partial_z \alpha-\partial_x \gamma=\frac{z+c}{(y-b)^2+(z+c)^2}-\frac{z-c}{(y+b)^2+(z-c)^2},\\
	w^*(\bfx) & = \partial_x\beta-\partial_y\alpha=\frac{x+a}{(z-c)^2+(x+a)^2}-\frac{x-a}{(z+c)^2+(x-a)^2} \\
\quad & \hspace{30mm} -\frac{y-b}{(y-b)^2+(z+c)^2}+\frac{y+b}{(y+b)^2+(z-c)^2},
\end{aligned}
\end{equation}
where $|\bfu^*| \rightarrow 0$ as $|\bfx| \rightarrow \infty$. These degenerated solutions are two dimensional in nature; for instance, $u^*$ does not necessarily decay at large $|y|$. So, we view them as degenerated `sources' or `sinks'.

The existence of bounded solutions is inconclusive. We propose a conjecture: The de-generated Laplace system (\ref{eeq}) admits only irrotational solutions on $\rr$. 

\subsection*{Upper half space}

In $\Upo_{U}: \{-\infty < x,y < \infty,\; 0 \leq z < \infty\}$, a convenient choice is given by
\begin{equation} \label{uhs}
\begin{split}
	\alpha & = \log\big(\; y^2+(z+c)^2\;\big)-\log\big(\; y^2+(z-c)^2\;\big)+ Ag_0,\\
	\beta & = \log\big(\; x^2+(z+c)^2\;\big)-\log\big(\; x^2+(z-c)^2\;\big)+ Ag_0,\\
	\gamma & = Ag_0.
	\end{split}
\end{equation}
Its curl is given by
\begin{equation}
\begin{aligned}
	u^* &= -\frac{2(z{+}c)}{x^2+(z{+}c)^2}+\frac{2(z{-}c)}{x^2+(z{-}c)^2},\\
	v^* & = \frac{2(z{+}c)}{y^2+(z{+}c)^2}-\frac{2(z{-}c)}{y^2+(z{-}c)^2},\\
	w^* & = \frac{2x}{x^2+(z{+}c)^2}-\frac{2x}{x^2+(z{-}c)^2}
-\frac{2y}{y^2+(z{+}c)^2}+\frac{2y}{y^2+(z{-}c)^2}.
\end{aligned}
\end{equation}
Specifically, $w^*|_{z=0}=0$, and $w^* \rightarrow 0$ as $z \rightarrow \infty$. These solutions are singular in the sense that, for instance, $u^*$ does not converge at finite $x$ or $z$, when $|y| \rightarrow \infty$.

\subsection*{An infinite layer}

In domain $\Upo_{L}: \{-\infty < x,y < \infty, 0 \leq z \leq L_z\}$,
method of images suggests that, in (\ref{uhs}), we replace $z$ by summations $z+2n_zL_z$ with integers $n_z$. The following expressions are derived from (\ref{eeq}):
\begin{equation} \label{ifl}
\begin{split}
	\alpha & = \log\big(\; y^2+(z+c)^2\;\big)-\log\big(\; y^2+(z-c)^2\;\big)+ Ag_0,\\
	\beta & = \log\big(\; x^2+(z+c)^2\;\big)-\log\big(\; x^2+(z-c)^2\;\big)+ Ag_0,\\
	\gamma & = Ag_0,
	\end{split}
\end{equation}
Let $z_{\pm}=z \pm c + 2 n_z L_z$. A solution for $\alpha$ is given by
\begin{equation*}
	\log\big(\; y^2+z_+^2\;\big)-\log\big(\; y^2+z_-^2\;\big).
\end{equation*}
Superposition of this solution gives $\alpha$ as
\begin{equation*}
	\sum_{n_z} \Big(\log\big(\; y^2 + (z {+} c {+} 2 n_z L_z)^2\;\big)-\log\big(\; y^2 + (z {-} c {+} 2 n_z L_z)^2\;\big) \Big)+Ag_0,
\end{equation*}
where the sum is taken over all values of $n_z$, $\sum_{n_z=-\infty}^{\infty}$. With a similar expression for $\beta$, we express $\bfu^*$ as follows:
\begin{equation} 
\begin{aligned}
	u^* &=\sum_{n_z} \bigg(\frac{2z_-}{x^2+z^2_-} - \frac{2z_+}{x^2+z^2_+}\bigg),\;\;\;
	v^* = \sum_{n_z} \bigg(\frac{2z_+}{y^2+z^2_+}-\frac{2z_-}{y^2+z^2_-}\bigg),\\
	w^* & = \sum_{n_z} \bigg( \frac{2x}{x^2+z^2_+}-\frac{2x}{x^2+z^2_-} -\frac{2y}{y^2+z^2_+}+\frac{2y}{y^2+z^2_-} \bigg),
\end{aligned}
\end{equation}
which shows $w^*|_{z=0}=w^*|_{z=L_z}=0$.

\subsection*{Finite rectangular parallelepipeds} 

Denote
\begin{equation} \label{drp}
	\Upo_P: {\;\;\;0\leq x\leq L_x,\;\;\;0\leq y\leq L_y,\;\;\;0\leq z\leq L_z}.
\end{equation}
Let $x_{\pm}=x \pm a + 2 n_x L_x$, $y_{\pm}=y \pm b + 2 n_y L_y$, where $n_x$ and $n_y$ are integers. A solution of the first equation in (\ref{eeq}) is given by
\begin{equation*} 
	\log\big(\;y^2_+ + z^2_+\;\big) - \log\big(\;y^2_+ + z^2_-\;\big) - \log\big(\;y^2_- + z^2_+\;\big) + \log\big(\;y^2_- + z^2_-\;\big) + Ax.
\end{equation*}
By cyclic permutation, the solutions for the second and the third equations are
\begin{equation*} 
	\log\big(\;z^2_+ + x^2_+\;\big) - \log\big(\;z^2_+ + x^2_-\;\big) - \log\big(\;z^2_- + x^2_+\;\big) + \log\big(\;z^2_- + x^2_-\;\big) + By,
\end{equation*}
and,
\begin{equation*} 
	\log\big(\;x^2_+ + y^2_+\;\big) - \log\big(\;x^2_+ + y^2_-\;\big) - \log\big(\;x^2_- + y^2_+\;\big) + \log\big(\;x^2_- + y^2_-\;\big) + Cz,
\end{equation*}
respectively. It follows that $\nabla{\cdot}\uppsi\neq 0$, provided $A{+}B{+}C \neq 0$. By the linearity property of system (\ref{eeq}), each of the solutions can be scaled. Denote the double sum
\begin{equation*}
	\sum_{j=-\infty}^{\infty}\;\sum_{k=-\infty}^{\infty} \;\;\;\;\;{\mbox{by}}\;\;\;\;\; \sum_{jk}.
\end{equation*} 
The $x$-component is found to be
\begin{equation} \label{r2}
\begin{aligned}
	u^*(\bfx) & = \sum_{n_x n_y}\bigg( \frac{y_+}{x^2_+ + y^2_+} - \frac{y_+}{x^2_- + y^2_+} - \frac{y_-}{x^2_+ + y^2_-} + \frac{y_-}{x^2_- + y^2_-}\bigg) \; - \\
	& \hspace{20mm}  \sum_{n_z n_x} \bigg(\frac{z_+}{z^2_+ + x^2_+}- \frac{z_+}{z^2_+ + x^2_-} - \frac{z_-}{z^2_- + x^2_+}+\frac{z_-}{z^2_- + x^2_-} \bigg).
	\end{aligned}
\end{equation}
The other formulas are found by cyclic permutation. In addition, the constraints, $\nabla{\cdot}\bfu^*=\nabla{\cdot}(\nabla{\times}\uppsi)=0$, and $\nabla{\times}\bfu^*=0$, are validated. The point is that the residue potentials always exist regardless of the scalar potential $\phi$.

So far, we have concentrated on finding closed-form solutions. For other simply connected domains where the boundaries are not described by regular geometry, we have to restore numerical methods to determine $\bfu^*$.

\section{Degeneracy in orthogonal curvilinear co-ordinates} \label{pfocc}
\subsection*{Cylindrical polar co-ordinates $(r,\theta,z)$}

Cylindrical polar co-ordinates are related to the Cartesian variables by
\begin{equation*}
	x=r \cos\theta,\;\;\; y = r \sin\theta,\;\;\; z=z,
\end{equation*}
where $r\geq 0$, and the polar angle $0 \leq \theta \leq 2 \pi$. A region in the interior or exterior of cylinders of finite axial length, $0 \leq z \leq L$, is simply-connected. The potential function is given by
\begin{equation}
	\frac{\partial^2 \phi}{\partial r^2}+ \frac{1}{r}\frac{\partial \phi}{\partial r} + \frac{1}{r^2}\frac{\partial^2 \phi}{\partial \theta^2}+ \frac{\partial^2 \phi}{\partial z^2}=0,
\end{equation}
subject to Neumann boundary condition.

To see if equation, $\nabla{\times}\nabla{\times}\uppsi=0$, admits solutions, we consider the divergence and the curl of the potential $\uppsi=(f,\;g,\;h)$
\begin{equation} 
	\nabla{\cdot}\uppsi=\frac{1}{r}\frac{\partial (rf)}{\partial r}+\frac{1}{r}\frac{\partial g}{\partial \theta}+ \frac{\partial h}{\partial z},
\end{equation}
and,
\begin{equation} 
\nabla\times \uppsi = \bigg\{ \frac{1}{r}\frac{\partial h}{\partial \theta}-\frac{\partial g}{\partial z},\;\;\;\;\;\frac{\partial f}{\partial z}-\frac{\partial h}{\partial r},\;\;\;\;\;\frac{1}{r}\bigg(\frac{\partial (rg)}{\partial r}-\frac{\partial f}{\partial \theta}\bigg) \bigg\},
\end{equation}
respectively. Direct calculation gives the analogous system to (\ref{eeq}):
\begin{equation} 
\left.
\begin{aligned}
\frac{1}{r^2}  \bigg(\frac{\partial^2 (rg)}{\partial \theta\partial r}-\frac{\partial^2 f}{\partial \theta^2}\bigg)-\frac{\partial^2 f}{\partial z^2}+\frac{\partial^2 h}{\partial z \partial r}&=0,\\
\frac{1}{r}  \frac{\partial^2 h}{\partial z \partial \theta}	- \frac{\partial^2 g}{\partial z^2}- \frac{\partial}{\partial r} \bigg(\frac{1}{r}\frac{\partial (rg)}{\partial r}\bigg)+\frac{\partial^2}{\partial r^2}\bigg(\frac{1}{r}\frac{\partial f}{\partial \theta}\bigg)&=0,\\
\frac{1}{r}\bigg(\frac{\partial f}{\partial z}-\frac{\partial h}{\partial r}\bigg)+ \frac{\partial^2 f}{\partial z \partial r} - \frac{\partial^2 h}{\partial r^2}-\frac{1}{r^2}\frac{\partial^2 h}{\partial \theta^2}+\frac{1}{r}\frac{\partial^2 g}{\partial \theta \partial z}&=0.
	\end{aligned}
\hspace{10mm} \right\}
\end{equation}
This system is seen to be over-determined. Nevertheless, it is straightforward to find some specific solutions. Intuitively, we observe a simple solution,
\begin{equation}
f=r/2,\;\;\;\;\; g = -\sigma z/(2 \pi r),\;\;\;\;\;h=1.
\end{equation}
Its curl has a line `source' or `sink' at every plane of constant $z$
\begin{equation} 
	\bigg(\;\frac{\sigma}{2 \pi r},\;\;\; 0,\;\;\; 0 \;\bigg);
\end{equation}
its divergence never vanishes. Alternatively, the same source solution is obtained from, $f=r/2,\;g=0,\;h=\sigma \theta/(2 \pi)$. In the source expression, $\sigma$ has the dimensions of planar flow rate 
[$\tt{m^2s^{-1}}$] per unit mass. 

Another set of solutions is given by
\begin{equation}
f=r/2,\;\;\;\;\; g = 1/r,\;\;\;\;\;h=-\cos(k\theta)/r^k,
\end{equation}
where $k$ is a positive integer. Velocity $\bfu^*$ is given by
\begin{equation} 
	\bigg(\;\frac{k\sin(k\theta)}{r^{k+1}},\;\;\;\;\; \frac{k\cos(k\theta)}{r^{k+1}},\;\;\;\;\; 0 \;\bigg).
\end{equation}
We can also choose $-k$, or replace the cosine function by $\sin(k\theta)$.

For the flow outside a cylinder of radius R, we combine the following two solutions,
\begin{equation} 
	f_1=f_2=r/2,\;g_1=g_2=1/r,\;\;h_1={-}r\cos\theta/(2 \pi R^2),\;\;h_2={-}\cos\theta/(2 \pi r),
\end{equation}
so that the curl is zero everywhere on cylinder's surface:
\begin{equation} 
	\bigg(\;\frac{\sin\theta}{2 \pi}\Big(\frac{1}{R^2}-\frac{1}{r^2}\Big),\;\;\;\;\; \frac{\cos\theta}{2 \pi}\Big(\frac{1}{R^2}-\frac{1}{r^2}\Big),\;\;\;\;\;0 \;\bigg).
\end{equation}
There is a constant radial flow far away from the cylinder. This absurd scenario is probably due to the geometry of the finite axial length where a symmetry in the $z$-direction does not exist.

There are several solutions where the curls are not zero the planes $z=0$, and $z=L$. The following set is one example
\begin{equation}
f=r/2,\;\;\;\;\; g = A (1/r-r/R^2)z,\;\;\;\;\;h=1,	
\end{equation}
where $A$ and $R$ are constants, and its curl is
\begin{equation}
	A\bigg(\;\frac{r}{R^2}-\frac{1}{r}\;\bigg),\;\;\;\;\;0,\;\;\;\;\; -\frac{2A}{R^2} z.
\end{equation}
Similarly, we derive a zero-curl solution,
$\big(\; r \theta (z^2{-}zL), \;r (z^2{-}zL)/2,\;r^2\theta(2z{-}L)/2 \;\big)$.

\subsection*{Spherical polar co-ordinates}

Spherical polar co-ordinates are such that
\begin{equation*}
	x=r \sin\theta \cos\phi,\;\;\; y = r \sin\theta \sin\phi,\;\;\; z=r \cos\theta,
\end{equation*}
where $r \geq 0$, the polar angle $0 \leq \theta \leq \pi$, and the azimuth angle $0 \leq \phi < 2 \pi$.
The potential is determined by boundary value problem, $\Delta_s\phi=0$, where the Laplacian operator is
\begin{equation} 
	\Delta_s=\frac{1}{r^2}\frac{\partial}{\partial r} \bigg(r^2 \frac{\partial}{\partial r}\bigg)+ \frac{1}{r^2\sin\theta}\frac{\partial}{\partial \theta}\bigg( \sin\theta \frac{\partial}{\partial \theta}\bigg) + \frac{1}{r^2 \sin^2\theta}\frac{\partial^2}{\partial \phi^2}.
\end{equation}
Denote $\uppsi=(f_{r},\; f_{\theta},\; f_{\phi})$. The components of the curl are
\begin{equation} 
	\frac{1}{r \sin\theta}\bigg( \frac{\partial (\sin\theta f_{\phi})}{\partial \theta}{-} \frac{\partial f_{\theta}}{\partial \phi}\bigg),\;\;\; \frac{1}{r}\bigg( \frac{1}{\sin\theta}\frac{\partial f_r}{\partial \phi}{-} \frac{\partial(r f_{\phi})}{\partial r}\bigg),\;\;\; \frac{1}{r}\bigg( \frac{\partial(r f_{\theta})}{\partial r}{-} \frac{\partial f_r}{\partial \theta}\bigg).
\end{equation}
We require that the divergence,
\begin{equation}
	\nabla{\cdot}\uppsi = \frac{1}{r^2}\frac{\partial}{\partial r}(r^2 f_r)+ \frac{1}{r \sin\theta}\frac{\partial}{\partial \theta} (\sin\theta f_{\theta})+\frac{1}{r \sin\theta} \frac{\partial f_{\phi}}{\partial \phi} \neq 0.
\end{equation}
The system of the degenerated Laplace equations now reads
\begin{equation} 
\left.
\begin{aligned}
\frac{\partial}{\partial \theta}\bigg( \sin\theta \frac{\partial(r f_{\theta})}{\partial r}\bigg)-\frac{\partial}{\partial \theta}\bigg( \sin\theta \frac{\partial f_r}{\partial \theta}\bigg) - \frac{1}{ \sin\theta}\frac{\partial^2 f_r}{\partial \phi^2}+\frac{\partial^2 (r f_{\phi})}{\partial \phi \partial r} &=0,\\
\frac{\partial^2(\sin\theta f_{\phi})}{\partial \phi \partial \theta}- \frac{\partial^2f_{\theta}}{\partial \phi^2}-r \sin^2\theta \frac{\partial^2(r f_{\theta})}{\partial r^2}+ r \sin^2\theta \frac{\partial^2 f_r}{\partial r \partial 
\theta}&=0,\\
\frac{1}{\sin\theta}\frac{\partial^2 f_r}{\partial r \partial \phi}-\frac{\partial^2 (rf_{\phi})}{\partial r^2}-\frac{\partial}{\partial \theta}\bigg( \frac{1}{\sin\theta}\frac{\partial (\sin\theta f_{\phi})}{\partial \theta}\bigg)+ \frac{\partial}{\partial \theta}\bigg( \frac{1}{\sin\theta}\frac{\partial f_{\theta}}{\partial \phi}\bigg)&=0.
	\end{aligned}
\hspace{3mm} \right\}
\end{equation}

Our task here is not to solve this system completely; we ought to be able to identify some particular solutions, no matter how simple they appear. By observation, we find one set of solutions,
\begin{equation} \label{s2}
	f_r=D/r,\;\;\;f_{\theta}=-A \phi \sin\theta /r + C/\sin\theta,\;\;\; f_{\phi} = -B/\sin\theta,
\end{equation}
for finite $A$, $B$, $C$ and $D$. Its curl is given by 
\begin{equation} \label{s5}
	\bfu^* = \bigg(\;\frac{A}{r^2},\;\;\;\;\;\frac{B}{r\sin\theta},\;\;\;\;\;\frac{C}{r \sin \theta}\;\bigg),
\end{equation}
that decays at large radii. The divergence equals to $(D - 2A \phi \cos\theta)/r^2$ which does not vanish for suitable $D$ and $A$.

Another set of solutions is found to be
\begin{equation} \label{s4}
	f_r=-A\phi\cos\theta/R + D/r,\;\;\;f_{\theta}=A \phi \sin\theta /R - C/\sin\theta,\;\;\; f_{\phi} = A/(R \sin\theta).
\end{equation}
Thus
\begin{equation} \label{s7}
	\bfu^* = \bigg(\;-\frac{A}{rR},\;\;\;\;\;-\frac{A}{rR\sin\theta}\big(\cos\theta+1\big),\;\;\;\;\;-\frac{C}{r \sin \theta}\;\bigg).
\end{equation}

The sum of solutions (\ref{s2}) and (\ref{s4}) gives a required velocity field with a zero radial component on radius $R$
\begin{equation} \label{s9}
	\bfu^* = \bigg(\;\frac{A}{r}\Big(\frac{1}{r}-\frac{1}{R}\Big),\;\;\;\;\;-\frac{A}{rR}\cot\theta,\;\;\;\;\;0\;\bigg).
\end{equation}
For bounded solutions, the above solutions are valid in well-chosen domains. At the unbounded points of elementary function $\cot\theta$, the singularities in the polar component in the vicinity of sphere's surface are integrable. If a spherical harmonics $\nabla\phi$ itself is singular, there are no technical reasons why this potential flow cannot be modified by adding multiples of velocities (\ref{s5}){-}(\ref{s9}). The essence is that a singularity in the theory of potential flow is a vague concept.

\section{Implication on Euler's equations}
\subsection{Potential flow and Euler's equations}

Consider inviscid flow-field $\bfu$, viz. $\nabla{\cdot}\bfu=0$, $\nabla{\times}\bfu=0$, and $\bfu=\nabla\phi+\nabla{\times}\uppsi$,
where $\Delta \phi = 0$, and $\nabla{\times}\nabla{\times}\uppsi=0$. Textbook formulation on potential flows, $\bfu=\nabla\phi$, represents an incomplete Helmholtz decomposition; the compatibility on the vector potential has never been treated in the technical literature. In practice, we have certain flexibility to resolve vector $\uppsi$: either $\nabla{\cdot}\uppsi=0$ or more generally $\nabla{\cdot}\uppsi\neq0$. The former case has been discussed in \S \ref{pflw} in some detail. In brief, Laplace equation $\Delta \uppsi=0$ has singular solutions at distinct locations, other than those of $\Delta \phi=0$. Strictly speaking, potential flows characterised in terms of sources, sinks or doublets, cannot be unique, though these singularities are instructive in visualising simplified flows. When the potential is non-solenoidal, either the solutions diverge at large distances, or they have trivial curls, as shown in \S \ref{dvps}. 

The Euler momentum equation,
\begin{equation}
\partial_t \bfu + (\nabla{\times}\bfu)\times \bfu = - \nabla(p/\rho+|\bfu|^2/2),
\end{equation}
can be re-written as
\begin{equation} \label{em4}
\nabla\bigg( \frac{\partial \phi}{\partial t} + \frac{p}{\rho} + \frac{1}{2}|\bfu|^2 \bigg) = -\nabla\times \frac{\partial \uppsi}{\partial t}.
\end{equation}
Since the components are orthogonal, we take divergence on (\ref{em4}) and get a generalised Bernoulli equation, 
\begin{equation}
\frac{\partial \phi}{\partial t} + \frac{p}{\rho} + \frac{1}{2}|\bfu|^2 = h(\bfx),\;\;\;\;\;\Delta h=0.
\end{equation}
For steady flows, we recover the Bernoulli principle
\begin{equation}
p+\rho(u^2+v^2+w^2)/2=C_h,
\end{equation}
where the same constant $C_h$ is taken on all streamlines. As expected, Bernoulli's law for the total pressure remains unchanged along well-defined streamlines, even though the velocity has its full orthogonal decomposition. In general, we must consider the boundary value problems for harmonic function $h$. If we pursue such a line of inquiry, we have to make extra effort to interpret its physical consequences.

\subsection{Potential flow in simply-connected domains}

In simply-connected domains $\Upo$ with $C^2$ boundary $\bdy$, the velocity potential is determined from the Neumann problem
\begin{equation}
	\Delta \phi = 0, \;\;\;\;\; \frac{\partial \phi}{\partial \vec{n}} \Big|_{\bdy}=\bfu\cdot {\vec {\bf n}}\big|_{\bdy} = \bfu_{s} \;(\neq 0),
\end{equation}
where $\bfu_{s}$ satisfies the usual compatibility condition. Moreover, the zero circulation leads to the determination of the non-solenoidal potential,
\begin{equation}
	\nabla{\times}\nabla{\times}\uppsi=0, \;\;\;\;\;(\nabla{\times}\uppsi) \cdot \vec{\bf n} \big|_{\bdy} = 0.
\end{equation}
The choice of homogeneous boundary condition is made, so that the potentials are independent of each other. The Euler momentum equation has an equivalent form,
\begin{equation} \label{eq2}
	\nabla{\cdot}\bfu=0,\;\;\;\;\; \nabla(p/\rho + |\bfu|^2/2)=\bfu \times (\nabla{\times}\bfu) = 0,
\end{equation}
subject to prescribed data $\bfu_s$. In terms of the potentials,
\begin{equation}
\bfu(\bfx)=\nabla \phi+\nabla{\times}\uppsi,\;\;\;\;\;p(\bfx)=-\frac{\rho}{2}|\bfu|^2+p_0.
\end{equation}
Note that the following is also an Euler solution for (\ref{eq2}),
\begin{equation}
\bfu'(\bfx)=\nabla \phi-\nabla{\times}\uppsi,\;\;\;\;\;p'(\bfx)=-\frac{\rho}{2}|\bfu'|^2+p_0,
\end{equation}
for identical data $\bfu_s$. In the finite rectangular parallelepipeds (\ref{drp}), solutions of $\uppsi$ have been found, see (\ref{r2}). Given $\bfu_s$, the solution of the Neumann problem for $\phi$ can be represented in terms of a Green function. In standard textbooks, the uniqueness proof for $\bfu$ assumes the complete absence of potential $\uppsi$. There is an underlying reason for the non-uniqueness: the principle of mass conservation alone is a weak control on the velocity field; the curl of the velocity must play a role even in irrotational flows. One simple example highlights the notion:
the constraint, $\nabla{\cdot}\tilde\bfu=0$, cannot uniquely specify the individual components of
\begin{equation*}
	\tilde\bfu(\bfx,t)=\big( u(\bfx,t)+Af(y,z,t), \; v(\bfx,t)+Bg(z,x,t),\; w(\bfx,t)+C h(x,y,t) \big),
\end{equation*}
for incompressible $\bfu$, and arbitrary bounded $f,g,h$. 

It is imperative to note that our discussion on potentials by no means invalidates any of the well-established physical laws in other branches of science. The inverse square law of Newton's universal gravitation is generalised by the idea of the gravitational potential. Likewise, Coulomb's law on the electric field due to charge density was established before the electric potential was introduced. Only {\it particular solutions} of the potential theory agree with these laws. According to Sommerfeld (\S19 of Sommerfeld, 1950), Newton never mentioned the gravitational potential. It was Lagrange who introduced the concept of potentials, see, for instance, Chapter III of Kellogg (1929). In fluid mechanics, it was Euler (\S 60, \S 66 and \S 67 of Euler, 1752) who first wrote about the velocity in terms of a (scalar) function, $\bfu=\nabla S$, and $S$ satisfies Laplace's equation. In the paper on the principles of fluid motion (1755), Euler reiterated the use of the potential in \S 26, and he did realise certain limitations of the potential in application (\S 29). Lagrange's paper (1781) was exclusively devoted to the use of the potential for irrotational flows in simply-connected domains. Apparently, the introduction of the potential simplifies computations; three components of velocity field are replaced by a single function. Helmholtz (1858) called this function the velocity potential (cf. his equation (1b)).

\subsection{Euler solutions in primitive variables}

Let $(\bfu,p)(\bfx,t>0)$ be a regular solution in the primitive variables, which is subject to perturbations at $t=t^*>0$. (A detailed analysis of formulation in terms of the primitive variables is given in Lam (2019).) For non-zero constant $\beta$, the combined flow is denoted by
\begin{equation} \label{euv}
	\bfu(\bfx,t) + \bfu'(\bfx) f(t),\;\;\;\; p(\bfx,t) + q(\bfx,t)f(t)/\beta,
\end{equation}
where $\nabla{\cdot}\bfu'{=}0$, and $f(t)$ is any continuous function of $t$; $f(t<t^*)=0$; $f(t\geq t^*)\neq0$. Substituting (\ref{euv}) into Euler's equations (\ref{euler}) and simplifying the result, we assert that the momentum is conserved, provided disturbed pressure $q$ satisfies Poisson's equation
\begin{equation} 
	\Delta q = - \rho \beta \;\nabla \cdot \Big( (\bfu{\cdot}\nabla)\bfu' + (\bfu'{\cdot}\nabla)\bfu + f(\bfu'{\cdot}\nabla)\bfu' \Big),
\end{equation}
at given time $t$. Thus, perturbation $\bfu'$ is frozen in inviscid flows. The choice for $\bfu'$ with finite energy is unlimited. One example is given by elementary functions
\begin{equation*} 
	(f,g,h)=\bigg( \frac{\sin(x{-}x_0 + q)}{\;(x{-}x_0)^{2n} + a^{2n}\;},\;\; \frac{\sin(y{-}y_0 + r)}{\;(y{-}y_0)^{2n} + b^{2n}\;},\;\; \frac{\sin(z{-}z_0 + s)}{\;(z{-}z_0)^{2n} + c^{2n}\;} \bigg),
\end{equation*}
which describe localised wave-packets (integer $n\geq1$), together with finite phase shifts, $q, r, s$, as well as non-zero stretching factors, $a,b,c$. A perturbation data is synthesised from them as
\begin{equation*} 
\bfu'(\bfx)=\big( A\, f g_y h_z,\;\; B\,g f_x  h_z,\;\; C\, h f_x g_y\big),
\end{equation*}
where $A+B+C=0$. 

In domains with regular boundaries $\Upo$, the normal perturbation velocity on $\bdy$ is zero in inviscid flows, 
\begin{equation} 
  \vec{n}\cdot (\bfu_s-\bfu')|_{\bdy}=0.
\end{equation}
The Neumann data for $q$ follow from
\begin{equation} 
	\nabla q = - \rho \beta \; \Big( (\bfu{\cdot}\nabla)\bfu' + (\bfu'{\cdot}\nabla)\bfu + f(\bfu'{\cdot}\nabla)\bfu' \Big),
\end{equation}
for known $\bfu|_{\bdy}$. Some examples are listed below.

\subsection*{Upper half space}

One particular perturbation is given by
\begin{equation*}
\begin{split}
u'(\bfx)=\;& \; 2y^3(1-2z^2)\exp(-x^2)\; \exp(-y^4)\; \exp(-z^2),\\
v'(\bfx)=\;& \; x(1-2z^2)\exp(-x^2)\; \exp(-y^4)\; \exp(-z^2),\\
w'(\bfx)=\;& \; 8xy^3 z\exp(-x^2)\; \exp(-y^4)\; \exp(-z^2),\\
\end{split}
\end{equation*}
where there is no perpendicular flow at the plane $z=0$. Both the $x-$axis and $y-$axis are homogeneous directions, any shift in $x \rightarrow x-x_0$ or $y \rightarrow y-y_0$ redefines the origin as $(x_0,y_0,0)$. Another solenoidal flow (odd integer $n \geq 1$) is
\begin{equation*} 
	\bfu'(\bfx)=\exp\big(-\alpha x^{n+1}-\beta y^{n+1}- \gamma z^{n+1}\big){\times }\big(A(yz)^n,\;B(zx)^n,\; C(xy)^n \big),
\end{equation*}
provided $\alpha A+\beta B+\gamma C=0$, $\alpha,\beta,\gamma > 0$, having non-zero boundary velocity
\begin{equation*}
	\bfu_s = C\exp\big(-\alpha x^{n+1}-\beta y^{n+1}\big)(xy)^n.
\end{equation*}

\subsection*{Rectangular parallelepipeds}

In domain (\ref{drp}), the following velocity field with no-flow $\bfu_s|_{\bdy}$ is incompressible 
\begin{equation} \label{vrect}
\begin{split}
u'(\bfx) & = A\; x(L_x-x)\; (L_y-2y)\; (L_z-2z),\\
v'(\bfx) & = B\; y(L_x-2x)\; (L_y-y)\; (L_z-2z),\\
w'(\bfx) & = C\; z(L_x-2x)\; (L_y-2y)\; (L_z-z),\\
\end{split}
\end{equation}
where $A+B+C=0$. Furthermore, vortex-dominated fluctuations, 
\begin{equation*}
\begin{split}
u'(\bfx)&= A \sin (2 \pi \alpha x) \cos (2 \pi \beta y) \cos (2 \pi \gamma z),\\
v'(\bfx)&= B \cos (2 \pi \alpha x) \sin (2 \pi \beta y) \cos (2 \pi \gamma z),\\
w'(\bfx)&= C \cos (2 \pi \alpha x) \cos (2 \pi \beta y) \sin (2 \pi \gamma z),\\
\end{split}
\end{equation*}
have zero through velocity on all walls, if
$(\alpha,\beta,\gamma)=(l/L_x, m/L_y, n/L_z)$, for integers, $l,m,n = 0,\pm1,\pm2,\pm3,\cdots$. The continuity demands $\alpha A + \beta B + \gamma C=0$. 
\subsection*{Arbitrary singularity}
The disturbed momentum, 
\begin{equation*}
	f'(t)\bfu' /f(t) +  (\bfu{\cdot}\nabla)\bfu' + (\bfu'{\cdot}\nabla)\bfu + f(t)(\bfu'{\cdot}\nabla)\bfu' = - \nabla q/(\rho \beta),
\end{equation*}
becomes unbounded or not, depending on the temporal function $f(t)$, which can be devised to generate singular perturbations, regardless of $\bfu'(\bfx)$. 
\section{Inviscid vorticity dynamics} \label{ivd}
The governing equation (\ref{vort}) on $\rr$ is displayed here 
\begin{equation*}
	\partial_t \upomega = (\upomega {\cdot} \nabla) \bfu  - (\bfu {\cdot} \nabla )\upomega,
\end{equation*}
which is subject to initial data $\upomega_0(\bfx)$. By the vorticity-velocity compatibility (\ref{cmp}), the velocity and its gradient are found from regular integrals,
\begin{equation*}
\bfu(\bfx) = -\frac{1}{4 \pi}\int_{\rr}\frac{\bfx-\bfx'}{|\bfx-\bfx'|^3}\times \upomega(\bfx') \; \rd \bfx',
\end{equation*}
and
\begin{equation*}
\nabla\bfu(\bfx) = \frac{1}{4 \pi}\int_{\rr}\nabla\Big(\frac{1}{|\bfx-\bfx'|}\Big)\nabla{\times}\upomega(\bfx') \; \rd \bfx',
\end{equation*}
respectively. The first integral confirms $\nabla{\cdot}\bfu=0$. In terms of a potential $\bfu=\nabla{\times}\uppsi$, an alternative for the velocity is to solve
\begin{equation} \label{p2}
\Delta \uppsi = - \upomega, \;\;\;\;\; \nabla{\cdot}\uppsi=c=\const
\end{equation}
We also assume that the vector potential decays, $\uppsi \rightarrow 0$ as $|\bfx| \rightarrow \infty$. The Biot-Savart law shows that 
\begin{equation} \label{p4}
\uppsi(\bfx) = \frac{1}{4 \pi}\int_{\rr} \frac{1}{|\bfx-\bfx'|}\;\upomega(\bfx')\; \rd \bfx'=\frac{1}{4 \pi}\int_{\rr} \frac{1}{\;|\bfx'|\;}\;\upomega(\bfx{+}\bfx')\; \rd \bfx'.
\end{equation}
Thus, the gauge constraint in (\ref{p2}) cannot be satisfied except in trivial flow. Denote the Fourier transform of vector ${\bf g}$ by
\begin{equation*}
	{\hat {\bf g}}(\bfk)=\frac{1}{\big(\sqrt{2 \pi}\big)^3}\int_{\rr}\exp( - \ri \;\bfk \cdot \bfx) \;{\bf g}(\bfx) \;\rd \bfx.
\end{equation*}
The Fourier transform of potential (\ref{p2}) yields
\begin{equation} \label{p6}
|\bfk|^2 \hat{\uppsi}(\bfk) = - \hat{\upomega}(\bfk)\;\;\; and \;\;\; \ri \bfk\cdot \hat{\uppsi}(\bfk)=c \delta(\bfk),
\end{equation}
where the choice of $\uppsi$-gauge is arbitrary ($c\neq0$). Evidently, the velocity and vorticity are incompatible: the vorticity has an absurd divergence in (\ref{p4}). In general, this integral miscalculates $\bfu$ and $\nabla\bfu$. Multiplying the first transformed equation of (\ref{p6}) by $\ri \bfk$ and applying the second, we derive an important result
\begin{equation*}
	\sum_{i,j=1}^3\big| - \bfk_i \bfk_j \hat{\uppsi}(\bfk)\big|^2 \neq \big|\hat{\upomega}(\bfk)\big|^2.
\end{equation*}
By solving (\ref{p2}), one is tempted to derive a gradient bound,
\begin{equation} \label{omg}
\| \nabla \bfu(\cdot,t)\|_{L^2} \leq c_0 \;\|\upomega(\cdot,t)\|_{L^2}.
\end{equation}
Note that both $\nabla\bfu$ and $\upomega$ are in $L^2$ or better. Intuition suggests that this bound cannot be true in general. Support that it holds for every constant $c_0$. The existence of this bound implies that there is a constant, $c_0^*$, such that
\begin{equation} \label{omg2} 
\| \upomega(\cdot,t)\|_{L^2} \leq c_0^* \; \|\nabla \bfu(\cdot,t)\|_{L^2}. 
\end{equation}
Let us recall that,
\begin{equation*} 
\begin{split}
\| \nabla \bfu\|^2_{L^2(\Upo)} & = \int_{\Upo} \Big( \frac{\partial u_i}{\partial x_j} \Big)^2 \rd \bfx, \;\;\;\mbox{and,}\\
\|\upomega\|^2_{L^2(\Upo)}&= \|\upomega\|^2_{L^2(\Upo)} + \int_{\Upo} (\nabla{\cdot}\bfu )^2 \rd \bfx\\
& =\| \nabla \bfu\|^2_{L^2(\Upo)} - 2 \int_{\Upo} \Big(\frac{\partial u_i}{\partial x_j} \Big) \Big(\frac{\partial u_j}{\partial x_i} \Big) \; \rd \bfx
+ 2 \int_{\Upo} \Big(\frac{\partial u_i}{\partial x_i} \Big) \Big(\frac{\partial u_j}{\partial x_j} \Big) \; \rd \bfx.
\end{split}
\end{equation*}
The last two integrals depend on the specification of the boundary velocity (such as, pre-assigned boundary strains or decays at infinity or periodicity). The point is that elliptic equation (\ref{p2}) is merely kinematic in nature. For evolution computations, the flow-field consistency, $\upomega = \nabla{\times}\bfu$, must hold at all times. Consider the case (\ref{vrect}) as an example. By direct calculation, we have
\begin{equation*}
\begin{split}
\|\nabla \bfu\|^2_{L^2}=(A^2+&B^2+C^2)L^3_x L^3_y L^3_z/27 
+ 2 A^2 L_x^5 L_y L_z (L_y^2 + L_z^2)/45 \\
& + 2 B^2 L_y^5 L_z L_x (L_z^2 + L_x^2)/45 
+ 2 C^2 L_z^5 L_x L_y (L_x^2 + L_y^2)/45,
\end{split}
\end{equation*}
and,
\begin{equation*}
\|\xi\|^2_{L^2}= 2L^3_x L_y L_z ( B^2 L^4_y /5 - BC L_y^2 L_z^2/3 + C^2 L_z^4/5 )/9.
\end{equation*}
The other two components, $\|\eta\|^2_{L^2}$ and $\|\zeta\|^2_{L^2}$, are obtained by cyclic permutations. Some numerical values are listed in table~\ref{tb1}. This `Euler flow' shows that $\|\upomega\|_{L^2}$ and $\|\nabla \bfu\|_{L^2}$ can be {\it a priori} bounded by one another, because each is finite. The gist is that bound (\ref{omg}) induces a circular argument. See also \ref{yud} below.

Our observation offers an alternative interpretation of Leray's energy relation (\ref{eiq}). Essentially, we are looking at $\|\bfu(\cdot,t)\|_{L^2(\rr)} \leq \|\bfu_0\|_{L^2(\rr)}$. The trouble is that the inequality sign causes conceptual difficulties, since it anticipates the occurrence of $\|\bfu_0\|_{L^2(\rr)} < c_1 \|\bfu(\cdot,t)\|_{L^2(\rr)} $. This incomprehension of two-way energy anomaly is entirely due to the logical fallacy just discussed. On the contrary, the conservation law must hold in inviscid flows.  {\it The equality sign} asserts the time-honoured principle and helps us rule out finite-time singularities. 
\begin{table} 
	\centering
\begin{tabular}{cccccccc} \hline \hline 
    $L_x$ & $L_y$ & $L_z$ & $A$ &  $B$ & $C$ &   $\|\upomega\|_{L^2}$ & 
    $\|\nabla\bfu\|_{L^2}$  \\ \hline \hline
    1 & 1 & 1 & 1 & 1 & -2 & 0.869 & 0.869   \\ 
	  1 & 4 & 0.25 & 2 & 2 & -4 & 2.763 & 7.224    \\ 
	  2 & 2 & 1 & 1 & 1 & -2 & 9.466 & 6.954    \\ \hline \hline
\end{tabular}
\caption{Field (\ref{vrect}) demonstrates $\| \nabla \bfu\|_{L^2} \leq c_0 \|\upomega\|_{L^2} \Longleftrightarrow \| \upomega\|_{L^2} \leq c_0^* \|\nabla \bfu\|_{L^2}$.} \label{tb1}
\end{table}

We conclude that equation (\ref{p2}) does not furnish bound (\ref{omg}) which misrepresents the strain fields over $t \in [0,T_*)$, where $T_*$ denotes a probable singular instant. Our derivation contradicts an estimate given in Beale {\it et al.} (1984). We must reappraise their analysis on establishing gradient $\|\nabla\bfu\|_{L^{\infty}}$ or inequality (21). The implication is that excessively large $\|\bfu\|_{L^{\infty}}$ can appear well prior to $T_*$. This potential break-down of the continuity may be accompanied by bounded vorticity. In the complete absence of viscous dissipation, the long-time accumulation,
\begin{equation*}
	\lim_{t \rightarrow \infty}\int_0^t \big\| \upomega (\cdot,s) \big\|_{L^{\infty}(\rr)} \; \rd s,
\end{equation*}
diverges and, potentially, instigates an artificial singularity for well-posed solutions.\footnote{Author's paper, {\ttfamily 1904.08385}, contains incomplete analyses and should be ignored altogether.} In a scenario of blow-up, one may opt to consider
\begin{equation*}
	\lim_{t \rightarrow T_*}\int_0^t \big\| \upomega (\cdot,s) \big\|_{L^{\infty}(\Upo)} \; \rd s \rightarrow \infty,
\end{equation*}
as a measure of vorticity growth, see, Luo \& Hou (2014), and Chen \& Hou (2022). This approach is very misleading unless other criteria are in place to monitor increases in the velocity. In spite of their numerical approximations, the expectation of a singular Euler solution in the incompressible framework is self-contradictory. 

Nevertheless, it is known that $\|\upomega(\bfx,t)\|_{L^{\infty}(\Upo_T)\times L^{\infty}(\rr)}$ is {\it a priori} bounded for $t \in \Upo_T=[\,0, T< \infty)$, see Lam (2019). As a consequence, we assert that finite-time blow-up never happens in incompressible Euler's flows, given the reality that there has been no convincing evidence to support the singularity hypothesis. It is perhaps a surprising fact that the all-important vorticity equation in three space dimensions, has not been formulated as the governing equation in the work of Leray (1934), Hopf (1951), and Ladyzhenskaya (1969), though the concept of vorticity has been introduced by Euler (1755), Lagrange (1781), Helmholtz (1895), as well as Lichtenstein (1929). In the paradigm of the primitive formulation (for real and ideal fluids), the perception of a singularity still misguides the regularity issue, even in the twenty-first century.  

\section{A postscript}

Laplace's equation (\ref{lp3}) for potential $\phi$ is linear; the velocity field on $\rr$ is necessarily ill-posed. In the case of source flow, for instance, the $x$-component can never be definite, as
\begin{equation*}
	u(\bfx)=\frac{1}{4 \pi}\lim_{N \rightarrow \infty} \sum_{i=1}^{N} m_i \; \frac{(x-x_i)}{r^3_i},\;\;\;\;\;r^2_i=|\bfx-\bfx_i| \leq R < \infty.
\end{equation*}
In fact, a velocity vector is given by Helmholtz's decomposition, $\bfu=\nabla\phi+\nabla{\times}\uppsi$, where there are two contributions. We omit an analogous theory in planes. It follows that potential flows cannot specify unique steady Euler solutions. 

Evidence to support singularity solutions of the {\it full} three-dimensional Euler equations is extremely limited. This is true even in the simplified flows where boundary effects have been removed. In retrospect, the disturbing state of affairs is that we are in serious doubt with what flow structures a credible singular field should resemble. In the absence of concrete examples and plausible rationales, the existence of the irregular blow-up solutions is by no means promising;
the postulation of fully developed turbulence as sequences driven by finite-time singularities remains unjustified within the framework of the continuum dynamics. 

The evolution of incompressible flow from given initial conditions proceeds over a series of mass-preserving processes instigated by the non-linearity, resulting in progressive flow states of numerous length scales of shears. The processes are dissipative where viscosity is the key; the energy conservation is fully observed. Specifically, energy dissipation means fluid elements or eddies lose their kinetic energy to fluid's internal energy until, ultimately, they are stationary. In view of thermodynamics, the energy of the small scales is not dissipated instantly so that the velocities of some eddies are persistently diminished while they are being pushed incessantly through the whole flow field by neighbouring, as well as distant stronger ones. Unavoidably, immobile fluid portions acquire vorticity due to shearing, or abruptly regain energy in the interaction as long as the surviving flow is strong enough, or there is a continuous supply of energy. 
It is not difficult to imagine that the motion of the multitudinous scales must be highly fluctuating and visually randomised, particularly in fluids having minute viscosity, where the dissipation well moderates, prolongs, and fragments over clustered regions. If we observe the motion at suitable fixed locations, i.e., in the Eulerian reference frame, the appearance of the dissipative scales is on-and-off, spasmodically over time, depending on the initial data. This natural phenomenon has a big name tag: intermittency. 

\appendix{Remarks on Yudovich's $L^p$-solutions} \label{yud}

Yudovich (1995) gave a revised account of his work on the well-posedness of unsteady Euler equations. Note that we are dealing with flows in two space dimensions. By default, velocity variable, 
$\bfu(\bfx,t)=(u,v)(\bfx,t)$; space variable, $\bfx=(x,y)$; and the non-vanishing vorticity $\zeta(\bfx,t)$ is everywhere normal to the $x{-}y$ plane. 
 
The initial-boundary value problem for the planar Euler equations is defined as
\begin{align} 
\nabla \cdot \bfu & = 0, && \text{(Continuity)} \\
\partial_t \bfu + (\bfu \cdot \nabla)\bfu &= - \nabla p, && \text{(Momentum)}\\
\bfu(\bfx,0) & = \bfu_0, && \text{(Initial data)} \\
\bfu\cdot \vec{n}|_{\bdy} & = 0, && \text{(Zero normal velocity)}
\end{align}
over domain $\Upo{\times}t \;(\in [0,\infty))$, where $\Upo \; ( \subseteq \real^2)$ has $C^2$ boundary $(\bdy)$. In unbounded domains, $\bfu_0$ is assumed to be smooth and compactly supported.

Yudovich (1995) claimed to have shown that there is a uniqueness theorem in certain functional spaces of incompressible flows (A1)-(A4) where the vorticity is singular or is $\in L^p(\Upo),\; p >1 $. In particular, his exposition depends on his Lemma 4.1 which states as follows: 
\begin{equation} \label{yd}
\| \nabla \bfu \|_{L^p(\Upo)} \; \leq \; C \; \frac{p^2}{p-1} \; \|\zeta \|_{L^p(\Upo)}\;\;\;\;\; (p > 1),
\end{equation}
where constant $C=C(\Upo)$. Note that this Lemma asserts that, at every fixed instant $t \in (0,\infty)$, {\it both the} $L^p$ {\it norms are bounded}, even though $\zeta$ may be singular at specific spatial locations $\bfx=\bfx_s$ (say).

The vorticity dynamics reads
\begin{align}
\partial_t \zeta + (\bfu \cdot \nabla)\zeta & = 0, && \text{(Angular Momentum)}\\
\zeta(\bfx,0) & = \nabla{\times}\bfu_0=\zeta_0, && \text{(Initial shear)}
\end{align}
where $\zeta=v_x-u_y$. There are {\it no} boundary conditions for the vorticity.

For arbitrary constant $C_0 > 0$ (where $C_0$ is assumed to have the correct dimensions), we have, at any given instant $t_* \in (0,\infty)$,
\begin{equation}
\begin{split}
\|\nabla \bfu\|_{L^p} & = \|\nabla \bfu\|_{L^p} + C_0 \; \int_{\Upo} \nabla {\cdot} \bfu \; \rd \bfx \\
\quad &  \leq \|\nabla \bfu\|_{L^p} + C_0 \; (\| u_x + v_y \|_{L^p}) \\
\quad & \leq \|\nabla \bfu\|_{L^p} + C_0 \; \big( \|u_x\|_{L^p} + \|v_y\|_{L^p} \big) \\
\quad & \leq \|\nabla \bfu\|_{L^p} + C_0 \; \big( \|u_x\|_{L^p} + \|v_y\|_{L^p} \big) + C_0 (\| |v_x - u_y| \|_{L^p}) \\
\quad & \leq \|\nabla \bfu\|_{L^p} + C_0 \; \big( \|u_x\|_{L^p} + \|v_y\|_{L^p} \big) + C_0 (\| |v_x| + |u_y| \|_{L^p}) \\
\quad & \leq \|\nabla \bfu\|_{L^p} + C_0 \; \big( \|u_x\|_{L^p} + \|v_y\|_{L^p} + \| v_x \|_{L^p} + \|u_y\|_{L^p} \big) \\
\quad & = (1+C_0) \; \|\nabla \bfu\|_{L^p},
\end{split}
\end{equation}
by virtue of the triangle inequality and Minkowski's inequality. For every $\Upo$ and $p$, there exists a $C_0$ (finite but sufficiently large) such that
\begin{equation*}
\|\zeta \|_{L^p(\Upo)} \; \leq \; (1+C_0) \; \|\nabla \bfu\|_{L^p(\Upo)}.
\end{equation*}

In other words, (\ref{yd}) inherently implies
\begin{equation} \label{ca}
\|\zeta \|_{L^p(\Upo)} \; \leq \; C_1 \; \|\nabla \bfu\|_{L^p(\Upo)},
\end{equation}
where $C_1=C_1(\Upo,p)$; Yudovich's Lemma 4.1 (1995) is a circular argument. 

The reason behind this logical fallacy is evident: the gradient bound, $\|\nabla \bfu\|_{L^p(\Upo)}$, contains the two derivatives, $u_x$ and $v_y$, of the continuity (where $u$ is differentiable in $x$; $v$ in $y$). But both are missing in $\|\zeta\|_{L^p(\Upo)}$. In applications, the absolute values of these two derivatives can become large compared to the gradients defining the vorticity, while the incompressibility is maintained over flow evolution.

Moreover, (\ref{yd}) is an elliptic estimate from $\bfu_{xx}+\bfu_{yy}=-\zeta$, taking into account of $\nabla{\cdot}\bfu=0$. Given the Neumann data, its solution is expressed in an integral relation, $\bfu = {\cal K} {*} \zeta$, see Yudovich's (4.1)-(4.4). Conversely, $\nabla {\cdot} {\cal K} \neq 0$. By a shift of the independent variables, we see that the incompressibility is valid as long as $\nabla{\cdot}\zeta=0$. But, if $\nabla{\cdot}\zeta$ vanishes everywhere, so does $\nabla{\cdot}(\zeta + C_2 x - C_2 y)$ for non-zero $C_2$, as there are no boundary conditions for the vorticity. In many numerical computations of initial-boundary value problem (A1)-(A4), it is a notorious issue that the velocity fields are often found to be non-divergence-free. The scenario is hardly surprising because the vorticity evolves, and is being updated by approximations.

For uniqueness, Yudovich (\S 3, 1995) used an energy estimate where the strict {\it inequality} suffers similar logical ambiguity. The essence is that the energy equality or the conservation law completely rules out Yudovich's singular solutions evolved from bounded initial energy. We assert that the existence of an analogous result to (\ref{ca}) in three space dimensions involves no additional principles. Thus, our exposition lends firm supports to the analysis of \S \ref{ivd}.

There are pitfalls in the stability theory of fluid motion (Lam, 2017). In spite of the rows over theory's foundation, there are attempts to generalise the concept of instabilities to Lebesgue's spaces, see, for instance, Friedlander \& Yudovich (1999). Such a generalisation invites another dimension of controversy, as it does not help resolve those inherent difficulties which are independent of disturbances' regularity. In fact, laboratory experiments do not validate the notion of the disturbance-driven laminar-turbulent transition, for a detailed discussion, see \S8 of Lam (2015). 

\appendix{Remarks on certain inequalities of analysis} \label{siq}
\subsection{The Sobolev inequality}

In functional analysis, Sobolev's inequality for scalar function $f$ is written as  
\begin{equation} \label{s0}
\| f \|_{L^q(\Upo)} \; \leq \; C \; \| \nabla f \|_{L^p(\Upo)},\;\;\;\;\;\; q=\frac{np}{n-p},
\end{equation}
where $n$ is the space dimension, and constant $C=C(n,p,\Upo)$. Function $f$ is assumed to be smooth $\Upo$. We assume that $\Upo \subset \real^n$ and has $C^2$ boundaries. In the case of $\real^n$, $f$ is compactly supported. 

Consider any incompressible velocity field $\bfu$ which is assumed to be smooth in domain $\Upo \subset \real^2$. In the case $\Upo=\real^2$, $\bfu$ is compactly supported. Its Lebesgue's norms ($ r\geq 1$) are denoted by
\begin{equation*}
\|\bfu\|_{L^r(\Upo)} = M_1 = M_1(\Upo,r) < \infty, \;\;\; \mbox{and} \;\;\; \|\nabla \bfu\|_{L^r(\Upo)} = M_2=M_2(\Upo,r) < \infty.
\end{equation*}
Consult books on functional analysis for further technical details, see, for instance, Adams \& Fournier (2003); McOwen (2002); Lieb \& Loss (2001); Beals (2004); Lawrence (1998).

In practice, we often deal with the initial-boundary value problem of the Euler (or the Navier-Stokes) equations, so that a fluid motion depends on both space and time, $\bfu=\bfu(\bfx,t)$. These space-wise $L^p$ norms are understood to hold at every given instant time $t \in [\:0,\:t^* < \infty)$. In the scenarios of a finite-time singularity, our discussion is valid over the time interval, when the motion is regular and possesses finite energy. 

In the light of (\ref{s0}), we see that
\begin{equation*}
\begin{split}
\|\bfu\|_{L^q} & = \|\bfu\|_{L^q} + C \int_{\Upo} \nabla {\cdot} \bfu \; \rd \bfx \\
\quad &  \leq \|\bfu\|_{L^q} + C \; (\| u_x + v_y \|_{L^p}) \\
\quad & \leq \|\bfu\|_{L^q} + C \; \big( \|u_x\|_{L^p} + \|v_y\|_{L^p} \big) \\
\quad & \leq \|\bfu\|_{L^q} + C \; \big( \|u_x\|_{L^p} + \|v_y\|_{L^p} \big) + C (\| |v_x - u_y| \|_{L^p}) \\
\quad & \leq \|\bfu\|_{L^q} + C \; \big( \|u_x\|_{L^p} + \|v_y\|_{L^p} \big) + C (\| |v_x| + |u_y| \|_{L^p}) \\
\quad & \leq \|\bfu\|_{L^q} + C \; \big( \|u_x\|_{L^p} + \|v_y\|_{L^p} + \| v_x \|_{L^p} + \|u_y\|_{L^p} \big) \\
\quad & = \|\bfu\|_{L^q} + C \; \|\nabla \bfu\|_{L^p}\\ 
\quad & \leq C_0\; \|\bfu\|_{L^q} + \|\bfu\|_{L^q},
\end{split}
\end{equation*}
where the last result has been established for sufficiently large constant $C_0\;(>0)$, if we {\it suppose} that relation (\ref{s0}) holds for the solenoidal $\bfu$. It follows that, for every given domain $\Upo$ and $p$, there exists a constant, such that
\begin{equation} \label{s1}
\| \nabla \bfu \|_{L^p(\Upo)} \; \leq \; C^* \; \| \bfu \|_{L^q(\Upo)}.
\end{equation}

{\it Conversely}, let (\ref{s1}) be true. Then its right hand side is bounded by the left, because 
\begin{equation*}
\begin{split}
\|\nabla \bfu\|_{L^p} & = \|\nabla \bfu\|_{L^p} + C^*_0 \; \int_{\Upo} \nabla {\cdot} \bfu \; \rd \bfx \\
\quad & \leq \|\nabla \bfu\|_{L^p} + C^*_0 \; \big( \|u_x\|_{L^p} + \|v_y\|_{L^p} \big) + C^*_0 (\| |v_x - u_y| \|_{L^p}) \\
\quad & \leq \; (1+C^*_0) \; \|\nabla \bfu\|_{L^p},
\end{split}
\end{equation*}
for a suitably chosen $C^*_0$. Thus, we obtain an analogous relation to (\ref{s0})
\begin{equation} \label{s0i}
\| \bfu \|_{L^q(\Upo)} \; \leq \; C_1 \; \| \nabla \bfu \|_{L^p(\Upo)},
\end{equation}
for $\nabla{\cdot}\bfu=0$, $\zeta=\nabla{\times}\bfu$, and $C_1= C_1( C)=C_1(n,p,\Upo)$.

The generalisation of the present procedures to three space dimensions is straightforward. We omit the details. 

It is an unfortunate fact that many claims on flow irregularity are the consequences of inequality (\ref{s0}). In $\rr$, the case $p=2$, i.e., the enstrophy, has been applied to incompressible velocity fields without justifications. See lemma 4.1 of Scheffer (1976). Thus, his idea of relating the weak turbulence solutions (Leray, 1934; Hopf 1951) to Hausdorff's dimension contains self-contradictory arguments. In a finite ball, the displayed inequality (2.9) of Caffarelli {\it et al.} (1982) is another example. Likewise, several applications of a modified version of (\ref{s0}) are found in Ladyzhenskaya and Seregin (1999). In the literature of instability proposal, there are undiscriminating uses of the inequality to manipulate the non-linear term, $(\bfu{\cdot}\nabla)\bfu$, for example, \S3 of Friedlander {\it et al.} (1997). 

\subsection{A note on Sobolev's embedding}

Although our attention in the present paper has been focused on fluid dynamics, we make an effort to explain the implications of inequality (\ref{s0}) on the embedding principle.

Let $f=f(x,y,z)$. Set
\begin{equation*}
\|f\|_{L^q(\rr)} = F_1 < \infty,\;\;\;\;\; \mbox{and} \;\;\;\;\; \|\nabla f\|_{L^p(\rr)}=F_2 < \infty,
\end{equation*}
for non-trivial real numbers $F_1, F_2$. By the properties of the real line, there exists a finite positive constant $C^* (\in \real)$ so that 
\begin{equation}
F_1 \; \leq \; C\; F_2\;\;\;\Longleftrightarrow \;\;\; F_2 \; \leq \; C^* \; F_1.
\end{equation}
These relations give rise to an embedding paradox. Let $W^{1,p}$ denote the Sobolev space. The embedding
\begin{equation} \label{seb1}
W^{1,p}(\rr)\; \hookrightarrow \; L^q(\rr),
\end{equation}
implies that
\begin{equation} \label{seb2}
L^q(\rr) \; \hookrightarrow \; W^{1,p}(\rr).
\end{equation}

Let $f$ be a locally-integrable function with no additional {\it a priori} smoothness. Consider the specific case $p=2$ and $q=6$. Embeddings (\ref{seb1}) and (\ref{seb2}) show that it is a matter of convenience to determine either $f \in L^6$ or $\nabla f \in L^2$. If $f$ is found to be regular in $L^6$, then the derivatives of $f$ must be square-integrable. Conversely, we can make efforts to bound $\nabla f$ as an $L^2$-function. In fact, we are dealing with an equivalent member of the Lebesgue space $L^6$. To improve the regularity, neither approach makes any progress, as space $L^6$ does not coincide with $W^{1,2}$, unless both spaces exclusively contain infinitely-differentiable functions.

The standard derivation of Sobolev's inequality is based on 
applying H{\"o}lder's inequality,
\begin{equation*}
\int_{\Upo} |f(\bfx)g(\bfx)|\; \rd \bfx \; \leq \; \|f\|_{L^{p'}(\Upo)}\; \|g\|_{L^{q'}(\Upo)},\;\;\; \Big(\; \frac{1}{p'}+\frac{1}{q'}=1,\;\;\; p',q'>1 \;\Big)
\end{equation*} 
and the fundamental theorem of calculus. Proofs given in textbooks are for a scalar function. See, for instance, Lawrence (1998) and McOwen (2002). For solenoidal vectors, the generalisation is by no means straightforward, if ever feasible. In addition, there is a fundamental issue which has long been neglected. Among many strategies to prove the H{\"o}lder inequality, we select the popular procedure which makes use of Young's inequality for product,
\begin{equation*}
A B \; \leq \; \frac{A^{p'}}{p'} + \frac{B^{q'}}{q'}.
\end{equation*}
This formula holds for positive $A$ and $B$. For every given set of $A$, $B$, $p'$ and $q'$, algebra asserts that there exists a positive constant $C'=C'(A,B,p',q')$ such that
\begin{equation*}
\frac{A^{p'}}{p'} + \frac{B^{q'}}{q'} \; \leq \; C'\: AB.
\end{equation*}
By analogy, we expect similar reversed inequalities involving Lebesgue's norms.

\subsection{The Ladyzhenskaya inequalities}

These inequalities were first derived in Chapter 1 of Ladyzhenskaya (1969); Foias {\it et al.} (2004); Ladyzhenskaya (1959). For any infinitely-differentiable scalar function $f$ with compact support in $\real^2$ or $\rr$, the following bounds hold 
\begin{equation} \label{lad2}
\| f \|_{L^4(\real^2)} \; \leq \; C_2 \; \big\| f \big\|^{1/2}_{L^2(\real^2)}\: \big\|\nabla f \big\|^{1/2}_{L^2(\real^2)},\;\;\;\;\;\;(n=2)
\end{equation}
and
\begin{equation} \label{lad3}
\| f \|_{L^4(\rr)} \; \leq \; C_3 \; \big\| f \big\|^{1/4}_{L^2(\rr)}\: \big\|\nabla f \big\|^{3/4}_{L^2(\rr)},\;\;\;\;\;\;(n=3)
\end{equation}
where $C_2$ and $C_3$ are constants. 

For the present purposes, there are no essential differences between the space dimensions. Let us consider any smooth two-dimensional solenoidal velocity $\bfu$. We have
\begin{equation*}
\begin{split}
\big\| \bfu \big\|^2_{L^4} & = \big\| \bfu \big\|^2_{L^4} + C^2_2 \; \| \bfu \|_{L^2} \; \int_{\real^2} (\nabla {\cdot} \bfu) \; \rd \bfx \\
\quad & \leq \big\| \bfu \big\|^2_{L^4} + C^2_2 \; \| \bfu \|_{L^2} \; \big( \; \|u_x\|_{L^2} + \|v_y\|_{L^2} + \| |v_x - u_y| \|_{L^2} \;\big) \\
\quad & \leq \; \big\| \bfu \big\|^2_{L^4} + C^2_2 \; \| \bfu \|_{L^2} \; \|\nabla \bfu\|_{L^2}, \\
\quad & \leq \; C'_2\;\big\| \bfu \big\|^2_{L^4} + \big\| \bfu \big\|^2_{L^4},
\end{split}
\end{equation*}
{\it if} (\ref{lad2}) is valid for $\bfu$. The (large) constant $C'_2$ is determined from the norms of $\bfu$. Thus, there exists a constant $C^*_2$ such that 
\begin{equation} \label{lad2i}
\big\| \bfu \big\|^{1/2}_{L^2(\real^2)}\: \big\|\nabla \bfu \big\|^{1/2}_{L^2(\real^2)} \; \leq \; C^*_2 \; \| \bfu \|_{L^4(\real^2)}.
\end{equation}
This bound is a direct consequence of (\ref{lad2}) where $f$ is replaced by $\bfu$. With minor modifications, our analysis is valid for finite domains ($\Upo \subset \real^2$) with $C^2$-boundary. 

For the three-dimensional case, it is convenient to start with
\begin{equation*}
\big\| \bfu \big\|^{4/3}_{L^4(\rr)} \; \leq \; C^{4/3}_3 \; \big\| \bfu \big\|^{1/3}_{L^2(\rr)}\; \|\nabla \bfu\|_{L^2(\rr)},
\end{equation*}
and, thus, to consider
\begin{equation*}
\big\| \bfu \big\|^{4/3}_{L^4(\rr)} = \big\| \bfu \big\|^{4/3}_{L^4(\rr)} + C^{4/3}_3 \; \big\| \bfu \big\|^{1/3}_{L^2{(\rr)}} \; \int_{\rr} (\nabla {\cdot} \bfu) \; \rd \bfx.
\end{equation*}

In conclusion, our ideas can be formulated as a criterion to re-appraise other well-known functional inequalities in dealing with divergence-free vector fields $\bfv$. The gist is the presence of a gradient norm $\|\nabla \bfv \|^s_{L^r(\Upo)}$.

\vspace{0.5cm}
\begin{acknowledgements}
\noindent 
15 June 2023

\noindent 
\texttt{f.lam11@yahoo.com}
\end{acknowledgements}
\end{document}